\documentclass[a4paper,scale=0.8,onecolumn,nobibnotes,nofootinbib]{revtex4}
\usepackage{enumitem}
\usepackage{amsmath,amssymb}
\usepackage{graphicx} 
\usepackage{epstopdf}
\usepackage{hyperref}
\usepackage{geometry}
\usepackage{subfigure}
\usepackage{makecell}
\usepackage{multirow}
\usepackage{float}
\usepackage{titlesec}
\usepackage{color}

\hypersetup{
    colorlinks=true,
    citecolor=blue,
    linkcolor=red,
    filecolor=magenta,
    urlcolor=cyan,}

\newcommand{\be}{\begin{equation}}
\newcommand{\ee}{\end{equation}}

\begin{document}

\title{Topology of nonlinearly charged black hole chemistry via massive gravity}
\author{Meng-Yao Zhang \footnote{gs.myzhang21@gzu.edu.cn}$^{1}$, Hao Chen\footnote{ haochen1249@yeah.net}$^{2}$, Hassan Hassanabadi\footnote{h.hasanabadi@shahroodut.ac.ir}$^{3,4}$,  Zheng-Wen Long\footnote{zwlong@gzu.edu.cn (Corresponding author)}$^{5}$, and Hui Yang\footnote{huiyang@gzu.edu.cn (Corresponding author)}$^{1}$}
\affiliation{$^1$ School of Mathematics and Statistics, Guizhou University, Guiyang, 550025, China.\\
$^{2}$ School of Physics and Electronic Science, Zunyi Normal University, Zunyi 563006, China.\\
$^{3}$ Faculty of Physics, Shahrood University of Technology, Shahrood, Iran.\\
$^{4}$ Department of Physics, University of Hradec Kr\'{a}lov\'{e}, Rokitansk\'{e}ho 62, 500 03 Hradec Kr\'{a}lov\'{e}, Czechia.\\
$^{5}$ College of Physics, Guizhou University, Guiyang, 550025, China.
}

\date{\today}

\begin{abstract}
The classification of critical points of charged topological black holes (TBHs) in anti-de Sitter spacetime (AdS) under the Power Maxwell Invariant (PMI)-massive gravity is accomplished within the framework of black hole chemistry (BHC). Considering the grand canonical ensemble (GCE), we show that $d=4$ black hole have only one topological class, whereas $d\ge 5$ black holes belong to two different topology classes. Furthermore, the conventional critical point characterized by negative topological charge coincides with the maximum extreme point of temperature; and the novel critical point featuring opposite topological charge corresponds to the minimum extreme point of temperature. With increasing pressure, new phases emerge at the novel critical point while disappear from the conventional one. Moreover, a atypical van der Waals (vdW) behavior is found in $d\ge 6$ dimensions, and the anomaly disappears at the traditional critical point. In the limit of nonlinearity parameter $s\to1$, different topology classes are only obtained in the GCE and they may not exist within the canonical ensemble. With the absence of electric potential $\Phi$, the neutral TBHs share the same topological classification results as the charged TBHs in the GCE of Maxwell-massive gravity.
\end{abstract}
\maketitle

\section{Introduction}
The existence of black holes mainly stems from theoretical derivation and experimental observation. From 1964 to 1970, Hawking and Penrose et al. proved a series of theorems about singularities \cite{P1,P2,P3,P4}. In 2015, the existence of gravitational waves was directly observed for the first time \cite{P5}, and in 2019, shadow of a black hole (BH) candidate was first observed with the Event Horizon telescope \cite{P6}. Both of these remarkable experiments indicate the possible existence of black holes. Particles entering the event horizon of a BH will not escape. Ever since physicists like Hawking and Bekenstein published major discoveries about black holes \cite{P7,P8,P9}, the thermodynamics of black holes became one of the most thrilling topics of gravitational theory. Phase transition is a highly common physical phenomenon in thermodynamic systems. Hawking and Page utilized anti-de Sitter (AdS) as the background and regarded Schwarzschild BH as a thermodynamic system. They revealed a phenomenon that when the BH temperature is low (event horizon is small), the Schwarzschild AdS BH is in a thermodynamically unstable state. As the temperature increases, the Schwarzschild BH has lower free energy and is in thermally stable state. That is, the stability of a BH relied on its critical temperature. This process is called the Hawking-Page (HP) phase transition \cite{SW3}. According to the AdS/CFT correspondence theory, the HP phase transition has garnered significant attention due in part to its potential implications within the context of gauge/gravity duality \cite{SW4}.

In recent years, the development of research on thermodynamic phase transitions in black holes has become increasingly intertwined with our approach to handling cosmological constants $\Lambda $. In the realm of theoretical physics, the concept of treating $\Lambda $ as a dynamic variable has been explored for decades. In fact, as early as 1984, Teitelboim and Henneaux proposed this idea in their research \cite{Y1,Y2}, which was followed by similar musings in the literature \cite{Y3,Y4}. However, it wasn't until 2009 that Kastor et al. made a groundbreaking discovery - they realized that $\Lambda $ could be considered as a variable thermodynamic pressure of a BH system \cite{Y5}. Recently, Mann et al. \cite{Y6} conducted an intresting study on the charged RN-AdS BH in four-dimensional spacetime and revealed there is a remarkable small/large BH phase transition occurs, which is strikingly similar to the gas-liquid phase transition observed in van der Waals (vdW) fluids. A new discovery \cite{Y7} has shown that BH systems and vdW fluids exhibit the same critical behavior, especially near critical points. Since then, additional types of phase transitions have been identified. For example, reentrant phase transitions (RPT) have been found in four-dimensional Born-Infeld-AdS (BI-AdS) black holes and high-dimensional single-spin Kerr-AdS \cite{Y8,Y9}, these black holes undergo a "LBH/SBH/LBH" phase transition over a certain pressure range. For more discussion about RPT, see \cite{Y10,Y11,Y12,Y13}. Triple points also occur in different types of black holes \cite{Y14,Y15,Y16,Y17}. Additionally, The $\lambda $-line BH phase transition \cite{Y18} has been presented in cubic Lovelock gravity.

 The similarity between the specific holographic phase transition and its real-world counterparts warrants further investigation. Black hole chemistry (BHC) has attracted the interest of many researchers \cite{SW11,SW12}. In general, the study of BHC is usually considered in the thermodynamic phase space, which is, the well-known extended phase space \cite{SW13}. However, the thermodynamic critical behavior can  still manifest even in non-extended phase spaces. For example, the phase transition of a charged topological Lovelock AdS BH is investigated in such a space \cite{SW14}. In the context of Einstein gravity, it has been observed that rotating or charged AdS black holes undergo phase transitions only in the canonical ensemble \cite{5,44}. However, modified theories of gravity have shown that this statement is no longer accurate. Research shows that the massive gravity could display distinct thermodynamic behavior from General Relativity (GR), and that graviton mass generate various new phase transitions for topological black holes (TBHs) \cite{D1}. Subsequently, the study of charged TBHs with massive gravity reveals a more intricate phase structure within the grand canonical ensemble (GCE) as opposed to the canonical ensemble \cite{SW15}. Therefore, exploring the behavior of black hole in a modified gravitational environment offers greater potential for uncovering complex phase structures. In other words, modified gravities offer a suitable framework for examining BHC under various conditions. So the investigation of TBHs in more complex environments is particularly intriguing. The thermodynamic properties of charged AdS BH in cubic generalized quasi-topological gravity have been investigated \cite{bhc1}, revealing both familiar and novel critical behavior and phase transitions within the context of BHC.

 Massive gravity is a natural theory that modifies Einstein's GR, it endowed a non-zero mass term for spin-2 gravitons. One of the most intriguing aspects of this theory is that it could provide a possible description for the self-acceleration of our universe without considering dark energy or a bare cosmological constant. In general, a basic challenge to constructing theory of massive gravitational fields is the emergence of what is known as the Boulware-Deser (BD) ghost \cite{R1}. Fortunately, the later nonlinear theory of dRGT massive gravity \cite{R2,R3} eliminated the BD ghost by introducing appropriate interaction terms. Also, current experimental datas from LIGO indicate that the graviton mass was highly constrained \cite{R4,R5}. Until now, massive gravity provided us with numerous insights into the behavior and properties of black holes, including situations which involve black hole solutions \cite{R6,R7} and their thermodynamical properties, such as Ruppeiner geometry and microstructures \cite{R8,R9}. In addition, the vdW like behaviour of dRGT massive gravity black holes, as well as a rich variety of phase transitions and optical features like the shadow, were also studied \cite{D1,R10,R11,R12}. On the other hand, there is a good motivation for power-Maxwell (PM) nonlinear electrodynamics (NED), which has the potential to eliminate the singularity of the electric field of point-like charges \cite{D2}. The power Maxwell invariant (PMI) field as an interesting class of the NED source, it is significantly richer than that of the Maxwell field and has produced some interesting thermodynamics in the context of AdS. Based on the above motives, black hole solutions of the massive gravity in the presence of PMI electrodynamics are more interesting. In Ref. \cite{ba}, the phase transitions of nonlinearly charged TBHs with PMI-massive gravitons in the GCE are discussed. In addition to traditional phase transitions, an anomalous vdW behavior was observed.

Topology is an important mathematical tool, the topological arguments in the context of black hole physics can be traced back to \cite{R13,R14}. It is an interesting and important problem to distinguish between different types of black holes or the same type of black holes from a topological point of view and to draw general conclusions. Recently, Wei \cite{x1} proposed a new method to study the thermodynamic topological properties of black holes. This method is of great significance for the search for the potential characteristic properties of black holes and contributes to the in-depth understanding of quantum gravity theory. Subsequently, the topological exploration of black hole thermodynamics has set off an upsurge \cite{x2,x3,x4,x5,x6,x7}. One significant manifestation of this is the proposal \cite{sw1} regarding the topology of critical points associated with black holes. According to Duan's $\phi$-mapping theory \cite{m1}, each critical point within the framework can be associated with a topological charge. This charge corresponds to the winding number and can take either a positive or negative value. A critical point with a topological charge of -1 is commonly referred to as a traditional critical point, while one with an opposite topological charge is considered novel. Subsequently, a new classification scheme for critical points is proposed, wherein the traditional critical points become the points of annihilation and the new critical points become the points where new phases (stable and unstable) appear \cite{Ye1,Ye2}. The effectiveness of the scheme has been demonstrated in various critical points, and it also holds true for isolated critical points \cite{MB1}. The topological information of other types of black holes has been revealed, such as Lovelock AdS black holes \cite{NC1}, R-charged black holes \cite{NJ1}, and more examples \cite{MR1}. To uncover more intriguing topological information, we extend this topological method to the context of BHC. The investigation of critical points through the exploration of BHs in massive gravity presents an intriguing opportunity to gain a deeper understanding of their nature. The primary objective of this correspondence is to extend the topology of critical points to more intricate BH backgrounds.

The remaining sections of the paper are structured as follows: Sect. II provides a comprehensive review of the thermodynamically significant aspects of charged-AdS TBHs, which are obtained via PMI-massive gravity. Additionally, we briefly introduce the topological methods required for the third part.  In Sec. III, we categorize the critical points by considering the grand canonical ensembles. In Sec. IV, we elucidate the characteristics of the critical points and showcase the behavior of the isobaric curves. In Sec. V, part A presents the topology of the Maxwell-massive gravity system for the nonlinearity parameter $s\to1$. In part B, we discuss the topological classes of pure massive gravity system in no any ensemble. Remarks and conclusions are given in Sec. VI.
\section{Thermodynamics of charged-AdS TBHs in PMI-massive gravity and Topology} \label{sec3}
With the inclusion of a negative cosmological constant, our attention is directed towards the thermodynamics of $d$-dimensional charged topological black holes within the framework of PMI-massive gravity. The description of this AdS black hole solution is given by the following bulk action: \cite{ba}
\begin{equation}
\mathcal{I}_{\mathrm{b}}=-\frac{1}{16 \pi G_d} \int_{\mathcal{M}} d^d x \sqrt{-g}\left[R-2 \Lambda+m_g^2 \sum_{i=1}^{d-2} c_i \mathcal{U}_i(g, f)+\mathcal{L}_{\mathrm{PMI}}\right],
\end{equation}
in the given expression, $R$ represents the Ricci scalar. The cosmological constant $\Lambda$ is associated with the spacetime dimension $d$ through the relation $\Lambda=-\frac{(d-1)(d-2)}{2l^2}$, where $l$ corresponds to the AdS radius. The parameter $m_g$ denotes the graviton mass, while ${c_{i}}^{,}$s represent arbitrary constants representing the massive couplings. The Lagrangian density of PMI electrodynamics, denoted as $\mathcal{L}_\mathrm{PMI}$, can be expressed as $(-\mathcal{F})^s$, where $\mathcal{F}=F_{\mu \nu}F^{\mu \nu}$ and $F_{\mu \nu}=\partial_{[\mu} A_{\nu]}$ represents the Faraday tensor. Here, $s$ denotes the nonlinearity parameter. Further, ${\mathcal{U}_i}^{,}$s are interaction potentials which can be constructed from the $d\times d$ matrix $\mathcal{K}^\mu{ }_\nu=\sqrt{g^{\mu \alpha} f_{\alpha \nu}}$ and can be given as
\begin{equation}\label{SymmetricPolynomials}
\mathcal{U}_i (g, f)=\sum_{y=1}^i(-1)^{y+1} \frac{(i-1) !}{(i-y) !} \mathcal{U}_{i-y}\left[\mathcal{K}^y\right],
\end{equation}
here, $f$ represents a fixed symmetric tensor, referred to as the reference metric, which is given by
\begin{equation}\label{RefMetric}
	f_{\mu\nu} = \text{diag}(0,0,c_{0}^{2}h_{ij}),
\end{equation}
where $c_0$ is a positive constant. After applying the equations of the motion and considering the static black hole solution the ${\mathcal{U}_i}$ have the following form \cite{M1}
\begin{equation}
	\begin{split}
		\mathcal{U}_{1}&=[\mathcal{K}]=(d-2)\frac{c_{0}}{r},\\
		\mathcal{U}_{2}&=[\mathcal{K}]^{2}-[\mathcal{K}^{2}]=(d-2)(d-3)\frac{c_{0}^{2}}{r^{2}},\\
		\mathcal{U}_{3}&=[\mathcal{K}]^{3}-3[\mathcal{K}][\mathcal{K}^{2}]+2[\mathcal{K}^{3}]=(d-2)(d-3)(d-4)\frac{c_{0}^{3}}{r^{3}},\\
		\mathcal{U}_{4}&=[\mathcal{K}]^{4}-6[\mathcal{K}^{2}][\mathcal{K}]^{2}+8[\mathcal{K}^{3}][\mathcal{K}]+3[\mathcal{K}^{2}]^{2}-6[\mathcal{K}^{4}]=(d-2)(d-3)(d-4)(d-5)\frac{c_{0}^{4}}{r^{4}}.
	\end{split}
\end{equation}
Then we utilize the line element ansatz in $d=n+2$ dimensions for the dynamical metric $g_{\mu \nu}$
\begin{equation}
d s^2=-V(r) d t^2+\frac{d r^2}{V(r)}+r^2 h_{i j} d x_i d x_j, \quad(i, j=1,2,3, \ldots, n),
\end{equation}
the line element $h_{i j} d x_i d x_j$ takes on the form as follows

\begin{equation}
h_{i j} d x_i d x_j = d x_1^2+ \frac{\sin^2\left(\sqrt{k}x_{1}\right)}{k} \sum_{i=2}^{d-2} d x_i^2 \prod_{j-2}^{i-1} \sin^2 x_{j},
\end{equation}	
where $k=-1$, $0$ and $1$ correspond to the hyperbolic, planar and spherical horizon geometries and the constant curvature of them is $(d_2)(d_3)k$.

Throughout this paper, we will consistently use the notation $d_i=d-i$. The metric function $V(r)$ is derived as follows \cite{ba}
\begin{equation}
V(r)=k+\frac{r^2}{\ell^2}-\frac{m}{r^{d_3}}+m_g^2 \sum_{i=1}^{d_2}\left(\frac{c_0^i c_i}{d_2 r^{i-2}} \prod_{j=2}^i d_j\right)+\frac{2^s(2 s-1)^2 q^{2 s}}{d_2\left(d_1-2 s\right) r^{2\left(s d_4+1\right) /(2 s-1)}},
\end{equation}
the ADM mass of the BH is \cite{M1}
\begin{equation}
M=\frac{d_2 \omega  _{n}}{16\pi} m,
\end{equation}
where $\omega  _{n}$ denotes the volume of $n$-dimensional unit hypersphere, and pressure $P$ can be expressed as $P=-\Lambda /8$, within the extended phase space. The thermodynamic quantities of TBHs in grand canonical ensembles, including temperature $T$, entropy $S$, volume $V$, $U(1)$ charge and the Gibbs free energy $G$ can be determined by their horizon radius $r_+$ \cite{ba}
\begin{equation}
T=\left.\frac{1}{4 \pi} \frac{d V(r)}{d r}\right|_{r=r_{+}}=\frac{1}{4 \pi d_2 r_{+}}\left[d_2 d_3 k+\frac{d_1 d_2}{\ell^2} r_{+}^2+m_g^2 \sum_{i=1}^{d_2}\left(c_0^i c_i r_{+}^{2-i} \prod_{j=2}^{i+1} d_j\right)-\frac{2^s(d_1-2 s)^{2s}\Phi ^{2 s}}{(2s-1)^{2s-1} r_{+}^{2\left(s-1 \right) }} \right],
\end{equation}
\begin{equation}
S=\frac{\omega _{n}}{4} r_{+}^{d_2},
\end{equation}
\begin{equation}
V=\frac{\omega _n}{d_1} r_{+}^{d_1},
\end{equation}
\begin{equation}
Q =\frac{\omega _n2^{s-1}s}{4\pi}q^{2s-1},
\end{equation}
\begin{equation}
G_{\Phi }=H-TS-Q\Phi,
\end{equation}
here $H$ is the enthalpy of the system, and the electric potential $\Phi=\left(\frac{2 s-1}{d_1-2 s}\right) q r_{+}^{-\left(\frac{d_1-2 s}{2 s-1}\right)}$, the acceptable range of $s$ is $1\le s<\frac{d_1}{2}$ \cite{ba}.  In the limit $s\to1$, the Maxwell case was restored. Based on the proposal of assigning topological charge to critical points put forward in \cite{sw1}. The temperature of a BH is denoted as $T=T\left(S, P, x^i\right)$, subsequently, we utilize the following relation to derive the critical points
\begin{equation}
\left(\partial_S T\right)_{P, x^i}=0, \quad\left(\partial_{S,S} T\right)_{P, x^i}=0.
\end{equation}
Now, we obtain a new function of temperature by eliminating the thermodynamic pressure from eq. (14), and then construct a Duan's potential \cite{sw1}
\begin{equation}
\Psi=\frac{1}{\sin \theta} T\left(S, x^i\right) ,
\end{equation}
where $1/\sin \theta$ is an additional factor which facilitates the topological analysis. Basted on the Duan's $\phi$-mapping theory, a new vector field $\phi$ is defined as $\phi=(\phi^S,\phi^\theta)$, where
\begin{equation}
\phi ^S=(\partial _S \Psi )_{\theta ,x_i}, \phi ^\theta=(\partial _\theta \Psi )_{ S,x_i}.
\end{equation}
When $\theta=\frac{\pi }{2} $, the component $\phi ^\theta$ is always zero. A topological current can be described as follows \cite{m1}
\begin{equation}
J^\mu=\frac{1}{2\pi } \epsilon ^{\mu \nu \lambda }\epsilon _{ab}\partial_{ \nu }n^a\partial_{ \lambda  }n^b,
\end{equation}
where $\partial_{ \nu }=\frac{\partial}{\partial x^{\nu}}$ and $x^{\nu}=(t, r, \theta)$. The unit vector $n$ is given by $n=(n^1, n^2)$, where $n^1=\frac{\phi ^S}{\left \| \phi  \right \| }$, $n^2=\frac{\phi ^\theta }{\left \| \phi  \right \| } $. And the topological current satisfies $\partial_\mu J^\mu=0$. Finally, given a parameter region $\sum $, the corresponding topological charge is given by
\begin{equation}
\begin{aligned}
Q_t  =\int_{\sum }^{} j^0d^2x=\sum_{i=1}^{N}  w_i,
\end{aligned}
\end{equation}
where $w_i$ is the winding number for $\phi ^a(x^i)=0$. The topological charge $Q$ can be positive or negative at critical points, depending on whether the winding number is +1 or -1.  The total topological charge $Q_t$ of a thermodynamic system can be determined by summing up the charges at each critical point, which enables us to classify various thermodynamic systems based on their global properties. Based on the aforementioned approach, we shall investigate AdS charged black holes through PMI-massive gravity in varying dimensions.
\section{Topology of critical points}
Based on the discussion in section II, we will proceed to the computation of the thermodynamic function using equations (7), (9) and (15) for temperature $T$ as provided
\begin{equation}
\begin{aligned}
\Psi =\frac{1}{4 \pi  d_2 r_+\sin \theta }\left[ -\frac{ 2^{1+s} s \left(d_1-2 s\right)^{2 s}\Phi^{2 s}}{(2 s-1)^{2 s-1} } r_{+}^{2-2 s}+2 d_2 d_3 k_{eff} + m_g^2 d_2 c_0 c_1 r_{+}+m_g^2\sum_{i=3}^{d_2} \left(i c_0^i c_i r_{+}^{2-i}\prod_{j=2}^{i+1}d_j \right)\right],
\end{aligned}
\end{equation}
where $k_{eff}$ is the effective topological factor \cite{ba1}
\begin{equation}
k_{eff}\equiv [k+m_g^2c_0^2c_2].
\end{equation}
The vector field $\phi$ can be decomposed into its components, namely $\phi^{r_+}$ and $\phi^\theta$, which can then be computed from eq. (16) as follows
\begin{equation}
\phi^{r_+}=\frac{\csc \theta}{4 \pi  d_2 r_+^2} \left [\frac{ 2^{1+s} s \left(d_1-2 s\right)^{2 s}\Phi^{2 s}}{(2 s-1)^{2 s-2} } r_{+}^{2-2 s}-2 d_2 d_3 k_{eff} +m_g^2\sum_{i=3}^{d_2} \left(i(1-i) c_0^i c_i r_{+}^{2-i}\prod_{j=2}^{i+1}d_j \right) \right],
\end{equation}
\begin{equation}
\begin{aligned}
\phi^\theta =-\frac{\cot \theta \csc \theta }{4 \pi  d_2 r_+}\left[-\frac{ 2^{1+s} s \left(d_1-2 s\right)^{2 s}\Phi^{2 s}}{(2 s-1)^{2 s-1} } r_{+}^{2-2 s}+2 d_2 d_3 k_{eff} + m_g^2 d_2 c_0 c_1 r_{+}+m_g^2\sum_{i=3}^{d_2} \left(i c_0^i c_i r_{+}^{2-i}\prod_{j=2}^{i+1}d_j \right) \right].
\end{aligned}
\end{equation}
The normalized vector field can be obtained through $n=(\frac{\phi ^{r_+}}{\left \| \phi  \right \| },\frac{\phi ^\theta }{\left \| \phi  \right \| } )$. We then search for topological charges associated with critical points in different dimensions. Based on the topology, a topological charge $Q$ is assigned to each critical point. When a given contour $C$ is constructed enclosing a single critical point $CP$, its topological charge $Q$ is non-zero; conversely, if $C$ outsides the $CP$, its topological charge $Q$ is zero \cite{R6,R7,sw1}. When several critical points covered by a contour $C$, $Q$ will be the sum of the topological charge at each critical point. According to the method of calculating the topological charge at the critical point, the contours we construct are positive in the orthogonal plane. The critical point is located on the $\theta=\frac{\pi}{2}$ axis in the $(S, \theta)$ plane, which can be reparameterized to the $(r_+, \theta)$ plane using Eq. (10). For a critical point located at $(r_0, \frac{\pi}{2})$ we can write out
\begin{equation}
\left\{\begin{matrix}
r_+=a\cos\vartheta +r_0\\ \theta =b\sin\vartheta +\frac{\pi}{2},
\end{matrix}\right.
\end{equation}
here $\vartheta\in (0,2\pi )$. Along the contour $C$, we then calculate the deflection angle $\Omega (\vartheta )$ of the vector field $\phi$, it is defined as
\begin{equation}
\Omega (\vartheta )=\int_{0}^{\vartheta } \epsilon _{ab}n^a \partial_\vartheta n^bd\vartheta,
\end{equation}
one can compute the topological charge
\begin{equation}
Q=\frac{\Omega (2\pi)}{2\pi}.
\end{equation}

\begin{figure}[htbp]
	\centering
	\subfigure[]{
    \begin{minipage}[t]{0.45\linewidth}
		\centering
		\includegraphics[width=3in,height=3in]{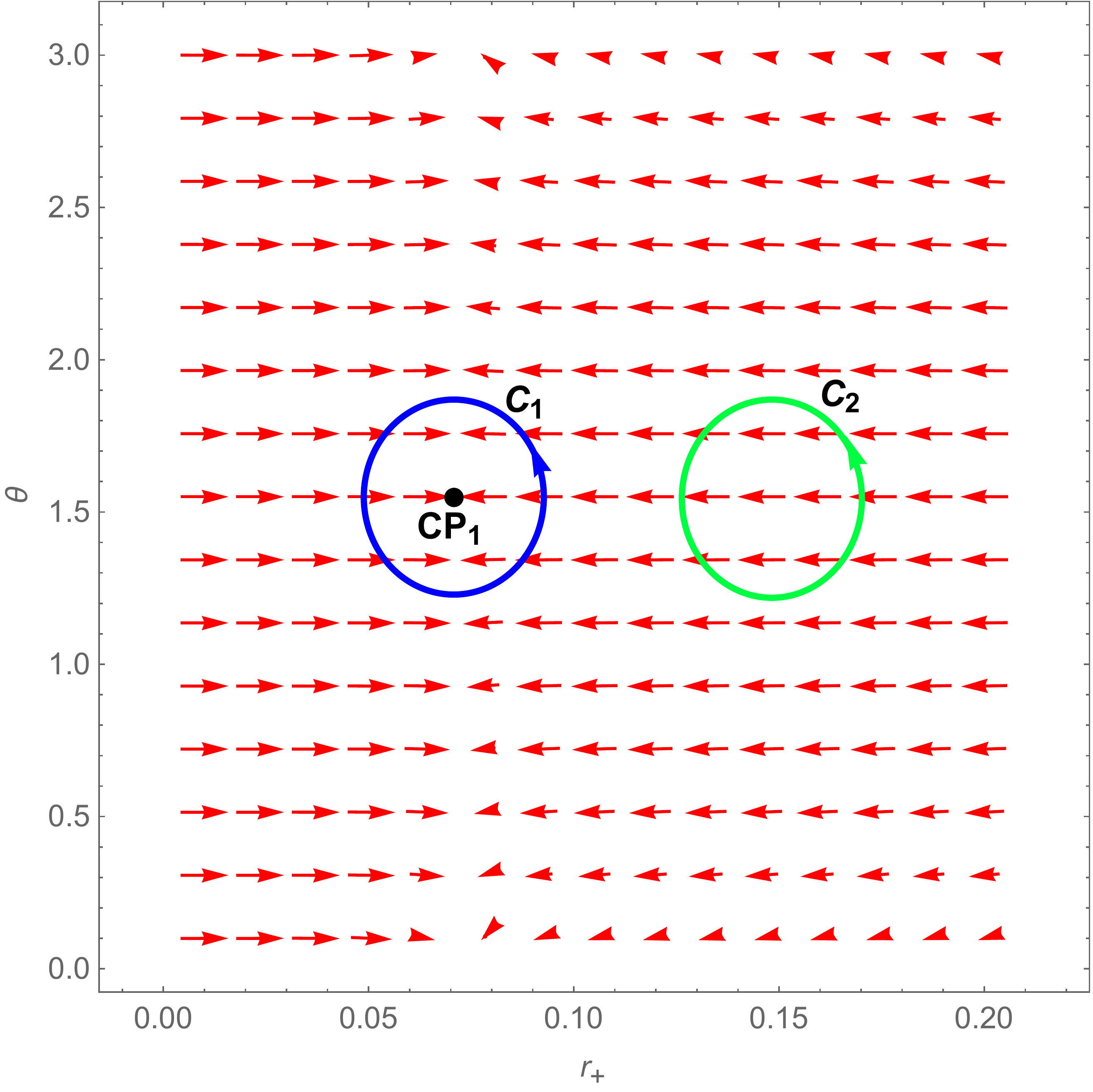}
		\end{minipage}%
            }%
    \subfigure[]{
    \begin{minipage}[t]{0.45\linewidth}
		\centering
		\includegraphics[width=2.6in,height=2in]{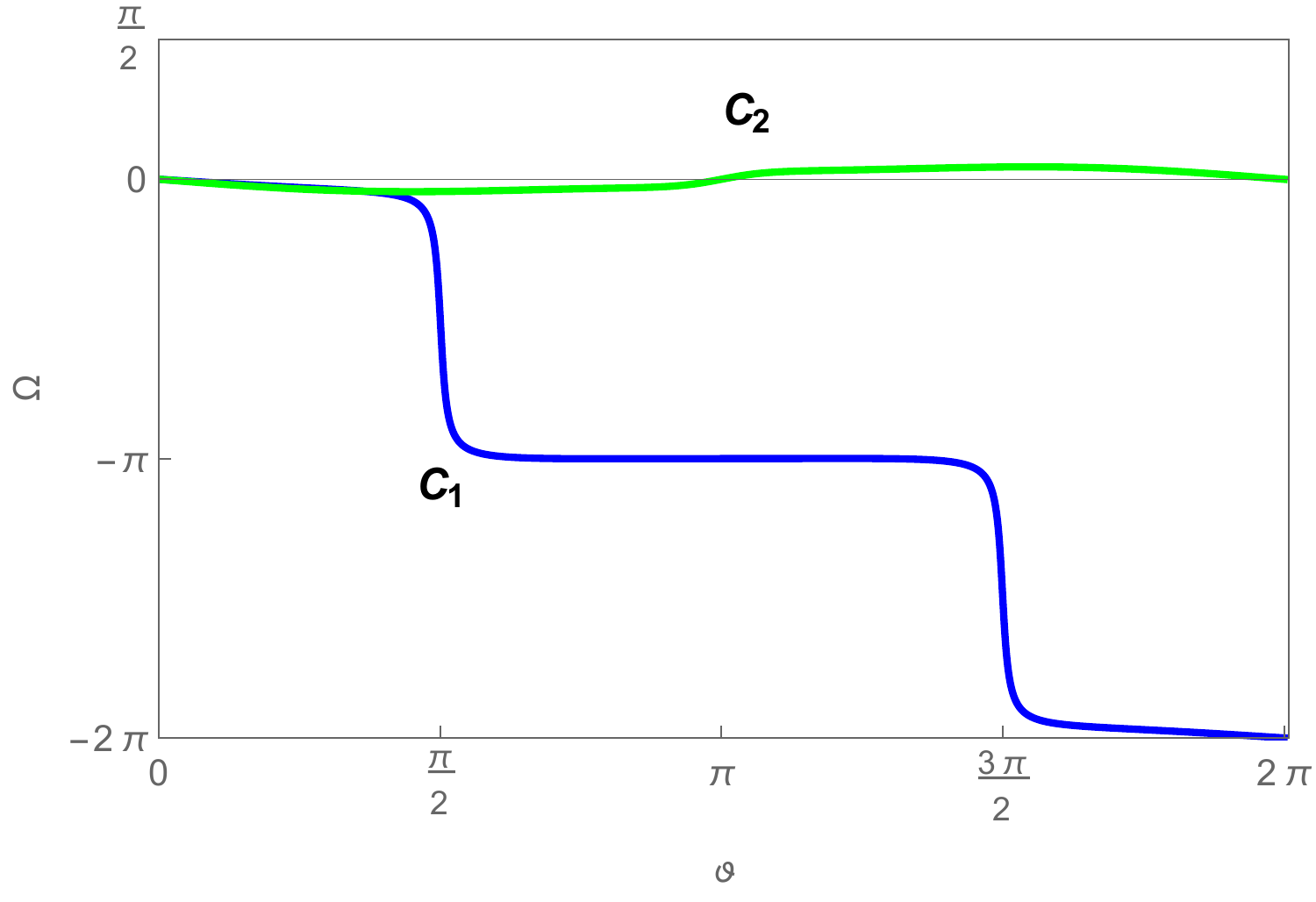}
		\end{minipage}%
            }%
     \centering
     \caption{$d=4$: (a) The red arrows represent the vector field $n$ for the charged TBHs in (PMI)-massive gravity in the GCE. The black dot located at $(r_+, \theta) =(0.07434, \pi/2 )$ which represents critical point $CP_1$. (b) The deflection angle $\Omega (\vartheta)$ as a function of $\vartheta$ for contours $C_1$ (blue curve) and  $C_2$ (green curve).}
\end{figure}
\begin{figure}[htbp]
	\centering
	\subfigure[]{
    \begin{minipage}[t]{0.45\linewidth}
		\centering
		\includegraphics[width=3in,height=3in]{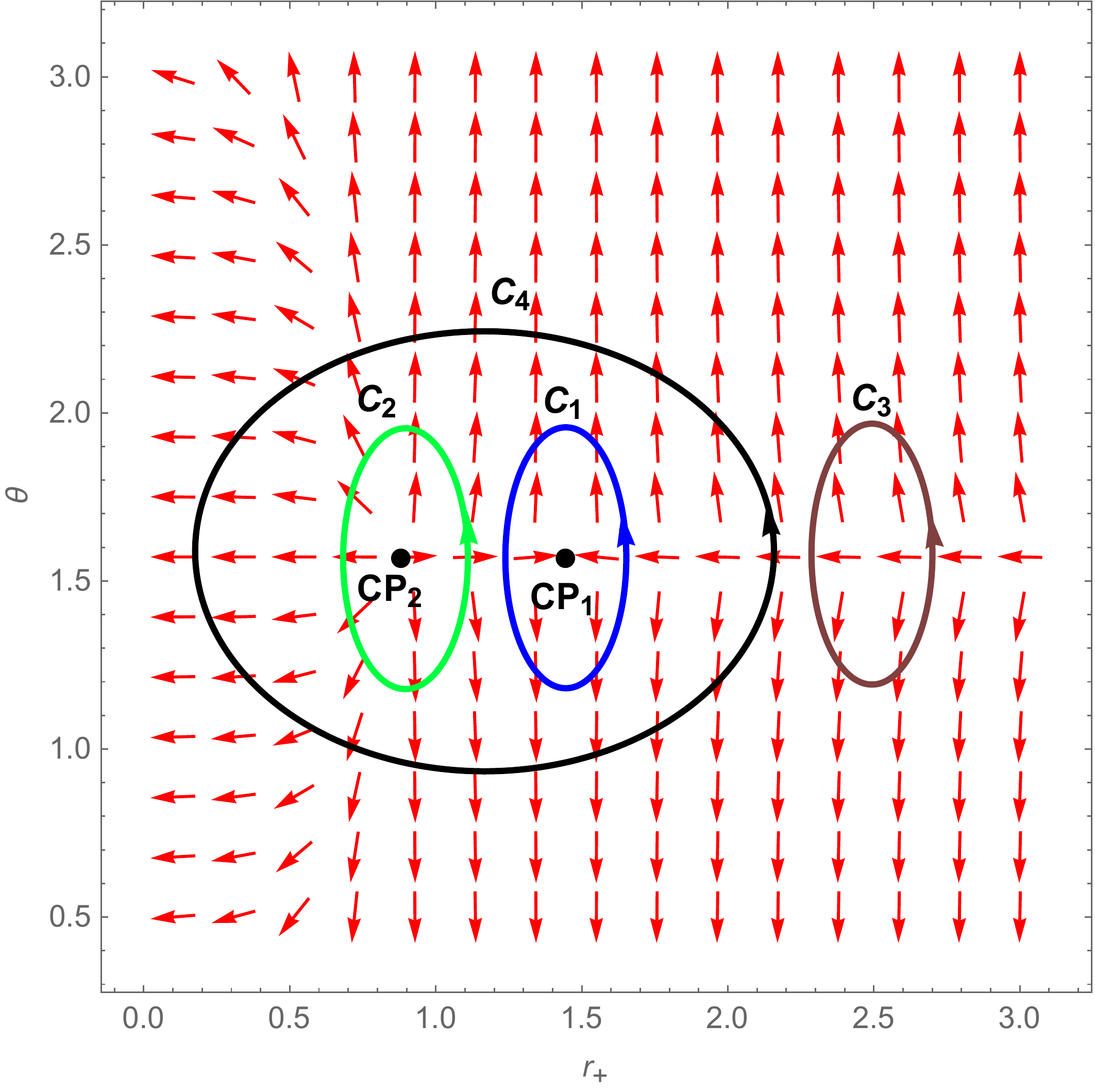}
		\end{minipage}%
            }%
    \subfigure[]{
    \begin{minipage}[t]{0.45\linewidth}
		\centering
		\includegraphics[width=2.6in,height=2in]{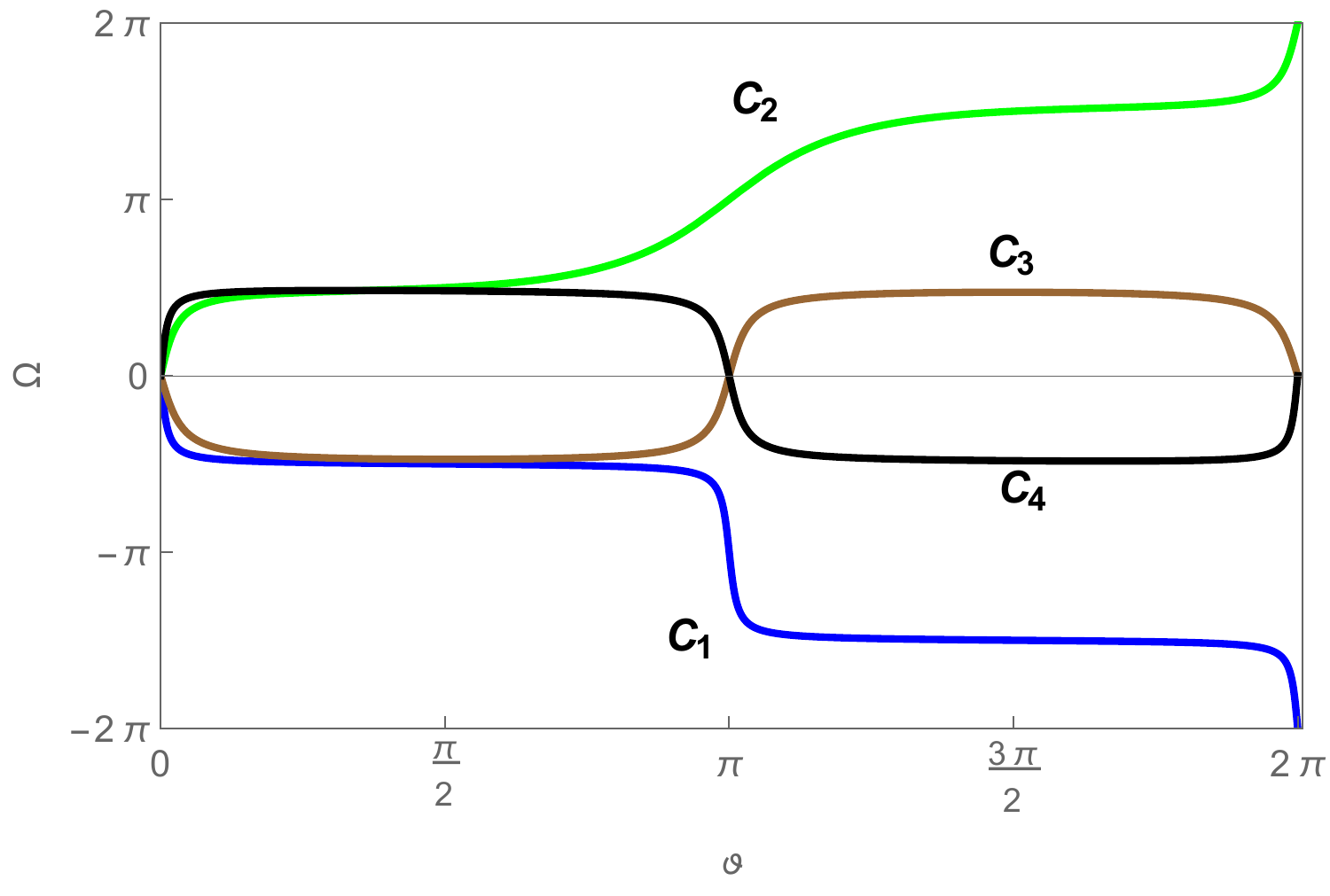}
		\end{minipage}%
            }%
     \centering
     \caption{$d=5$: (a) The red arrows represent the vector field $n$ for the charged TBHs in the GCE of PMI-massive gravity. The black dots are $(r_+, \theta) =(1.4510, \pi/2 )$ and $(r_+, \theta) =(0.89667, \pi/2 )$, which represent the critical points $CP_1$ and $CP_2$, respectively. (b) The deflection angle $\Omega (\vartheta)$ as a function of $\vartheta$ for contours $C_1$ (blue curve), $C_2$ (green curve), $C_3$ (brown curve) and $C_4$ (black curve).}
\end{figure}
\begin{table}[]
	\centering
	\caption{Critical values at $m_g=1$, $c_0=c_1=1$ for PMI-massive gravity system.}
\begin{tabular}{c|l|l|l|l}
 \hline \hline   $d$                                  &      & $CP_1$    & $CP_2$    & $CP_3$ \\
\hline                                             & $r_c$  & 0.07434 &  -      & -            \\
4($k_{eff}=1, \Phi=1, s=1.2$)                        & $T_c$  & 0.69123 &  -      & -            \\
                                                   & $P_c$  & 1.19981 &  -      & -         \\
\hline                                             & $r_c$  & 1.45104 & 0.89666 & -            \\
5($k_{eff}=0.7, c_3=0.4, \Phi=1.3, s=1.3$)           & $T_c$  & 0.11448 & 0.11338 &  -           \\
                                                   & $P_c$  & 0.00435 & 0.00360 &-            \\
\hline                                             & $r_c$  & 0.77590 & 2.35883 & 1.21566     \\
6($k_{eff}=0.9, c_3=-0.7, c_4=0.6, \Phi=7, s=2.3$)   & $T_c$  & 0.14820 & 0.14561 & 0.13963      \\
                                                   & $P_c$  & 0.01327 & 0.00773 & 0.00408      \\
 \hline                                            & $r_c$  & 0.24091 & 2.65128 & 1.39651     \\
6($k_{eff}=0.9, c_3=-0.7, c_4=0.6, \Phi=4.5, s=2.2$) & $T_c$  & 3.86646 & 0.14249 & 0.13458     \\
                                                   & $P_c$  & 10.0013 & 0.00684 & 0.00254     \\
\hline \hline
\end{tabular}
\end{table}
\begin{table}[]
	\centering
	\caption{Contours at $m_g=1$, $c_0=c_1=1$ for PMI-massive gravity system.}
\begin{tabular}{c|l|l|l|l|l}
 \hline \hline $d$ & & $C_1$ & $C_2$ & $C_3$ & $C_4$ \\
\hline                                             & $a$    & 0.07    &  0.07      & -           &- \\
4($k_{eff}=1, \Phi=1, s=1.2$)                        & $b$    & 0.4     &  0.4       & -           &- \\
                                                   & $r_0$  & 0.07434 &  0.15      &  -          &-\\
\hline                                             & $a$    & 0.15    & 0.15       & 0.15        &0.15\\
5($k_{eff}=0.7, c_3=0.4, \Phi=1.3, s=1.3$)           & $b$    & 0.4     & 0.4        & 0.4         &0.4\\
                                                   & $r_0$  & 1.45104 & 0.89667    & 2.4        &1.17 \\
\hline                                             & $a$    & 0.15    & 0.15       & 0.15        &1.1\\
6($k_{eff}=0.9, c_3=-0.7, c_4=0.6, \Phi=7, s=2.3$)   & $b$    & 0.4     & 0.4        & 0.4         &0.5\\
                                                   & $r_0$  & 0.77590 & 2.35883    & 1.21566     &1.57\\
 \hline                                            & $a$    & 0.15    & 0.15       & 0.15        &-\\
6($k_{eff}=0.9, c_3=-0.7, c_4=0.6, \Phi=4.5, s=2.2$) &$b$   & 0.4     & 0.4        & 0.4         &-\\
                                                   & $r_0$  & 0.24091 & 2.65128    & 1.39651     &-\\
\hline \hline
\end{tabular}
\end{table}
\subsection{$d=4$}
We can find one critical point in $d\ge4$ dimension if $s\ne1$ ($s\to 1$ was discussed in Sec. V). When $d=4$, there are only the first two massive couplings $(c_1)$ and $(c_2)$ are present in thermodynamic function (19), we simply assume they are non-zero. Refer to TABLE I for the critical value in this instance. Reference \cite{ba} displays the a typical swallowtail behavior, indicating the occurrence of a SBH/LBH phase transition. The behavior of $\theta$ with the radius $r_+$ for the vector field $n$ is represented in Fig. 1(a), which reveals that there exists a single critical point located at $(r_+, \theta) =(0.07434, \pi/2 )$. We construct two contours $C_1$ and $C_2$, $C_1$ encloses the critical point $CP_1$, $C_2$ doesn't enclose any critical point (see TABLE II for the parameters of contours). In Fig. 1(b), the results of the deflection angle $\Omega (\vartheta)$ are presented along the contours $C_1$ and $C_2$, we find that $\Omega (\vartheta)=-2\pi$ and $0$, for $C_1$ and $C_2$, respectively, when $\vartheta=2\pi$. Then, the topological charge of critical point $CP_1$ is given by $Q_{CP_1}=-1$.  While contour $C_2$ does not enclose any critical point and therefore has a topological charge of zero. Based on the classification in \cite{sw1}, $CP_1$ can now be considered a conventional critical point. Therefore, the total topological charge in this case is $Q=Q_{CP_1}=-1$. When considering higher-dimensional spacetime, the most direct approach to obtain one critical point is by assuming the absence of other massive couplings while maintaining non-zero values for $c_1$ and $c_2$. However, it is important to acknowledge that fine-tuning other relevant parameters also allows us to find one (physical) critical point.
\begin{figure}[htbp]
	\centering
	\subfigure[]{
    \begin{minipage}[t]{0.45\linewidth}
		\centering
		\includegraphics[width=3in,height=3in]{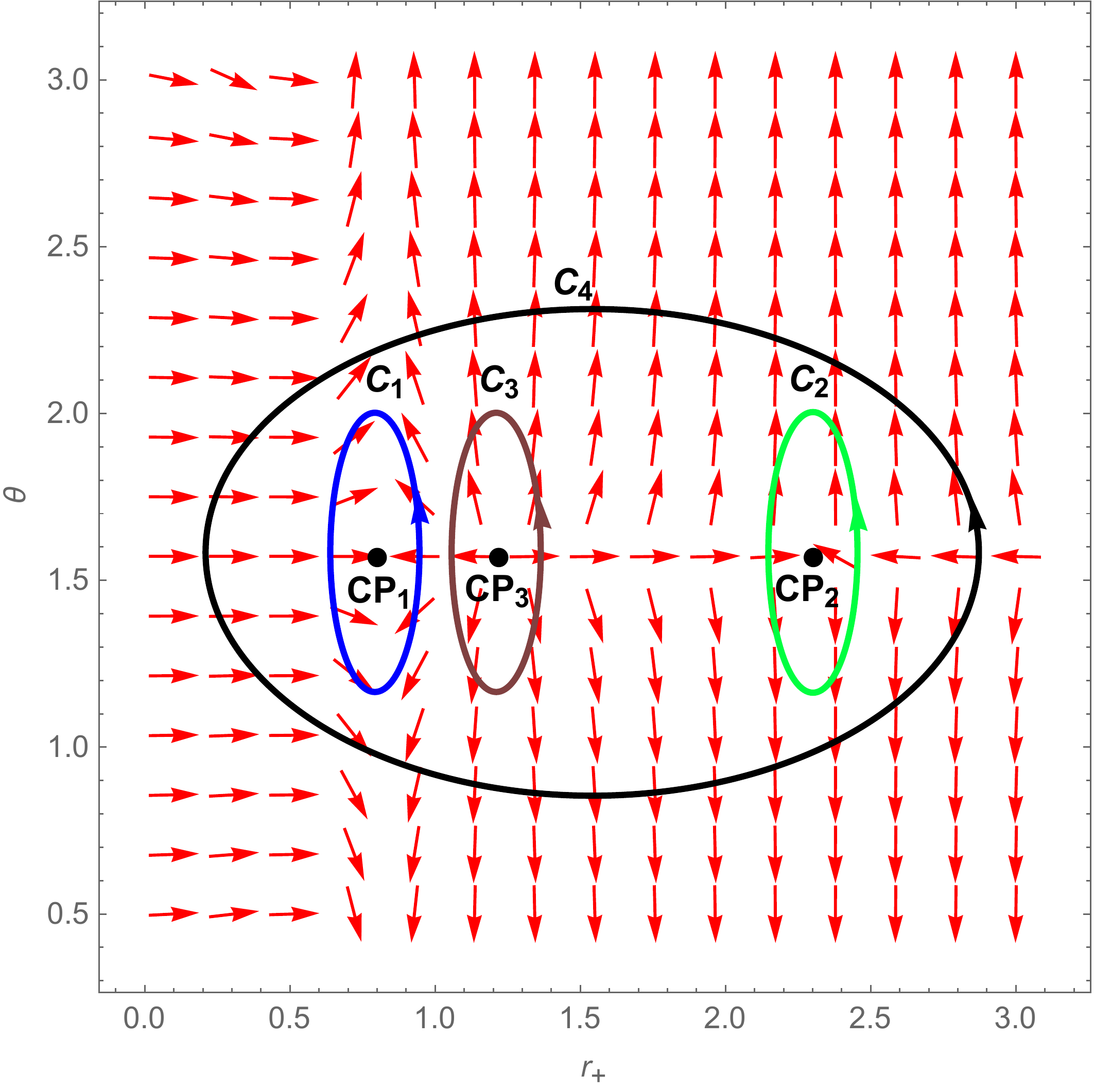}
		\end{minipage}%
            }%
    \subfigure[]{
    \begin{minipage}[t]{0.45\linewidth}
		\centering
		\includegraphics[width=2.6in,height=2in]{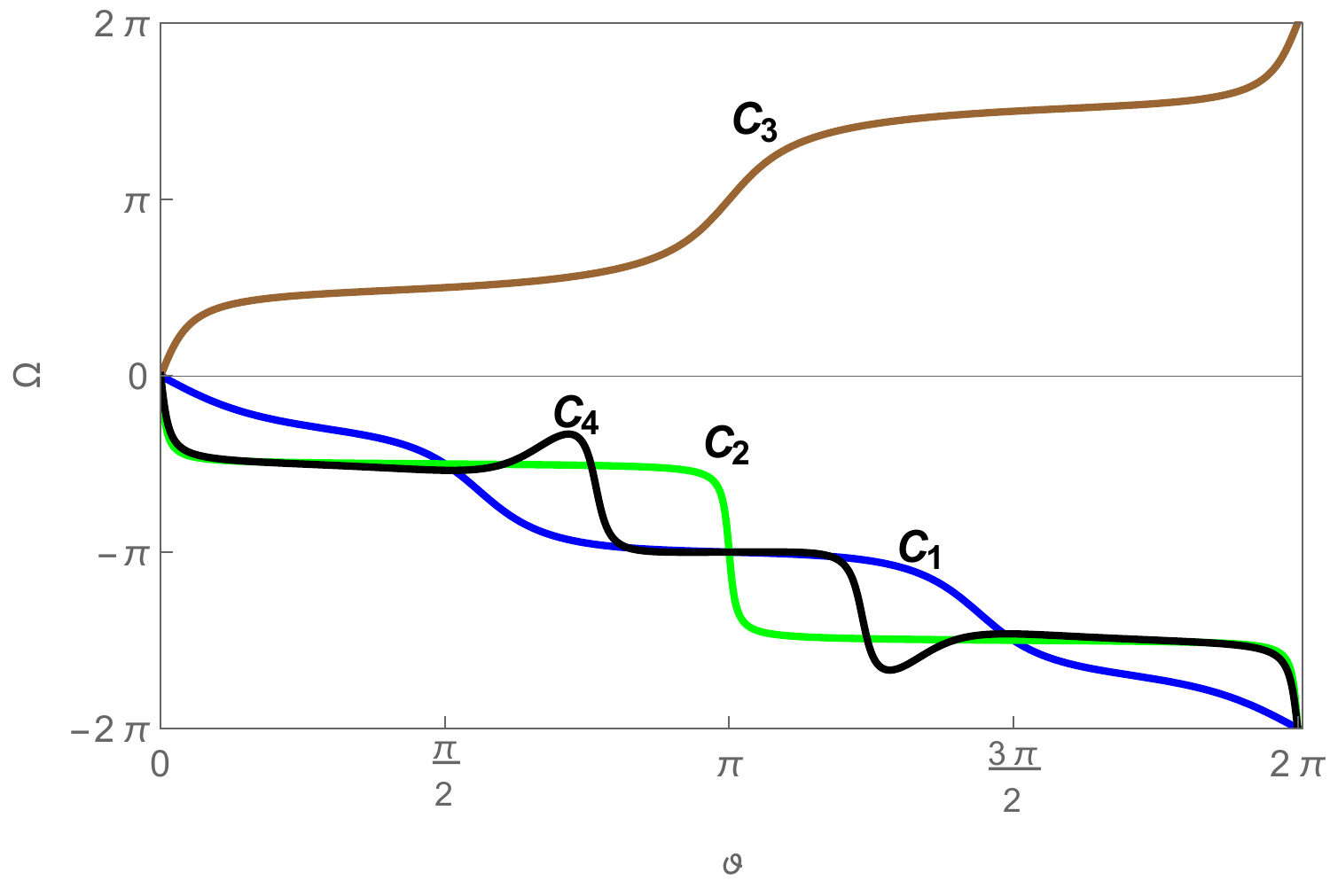}
		\end{minipage}%
            }%
     \centering
     \caption{$d=6$(case 1): (a) The red arrows represent the vector field $n$ for the charged TBHs in the GCE of PMI-massive gravity. The black dots are $(r_+, \theta) =(0.7759, \pi/2 )$, $(r_+, \theta) =(2.3588, \pi/2 )$ and $(r_+, \theta) =(1.2157, \pi/2 )$ which represent the critical points $CP_1$, $CP_2$ and $CP_3$, respectively. (b) The deflection angle $\Omega (\vartheta)$ as a function of $\vartheta$ for contours $C_1$ (blue curve), $C_2$ (green curve), $C_3$ (brown curve) and $C_4$ (black curve).}
\end{figure}
\begin{figure}[htbp]
	\centering
	\subfigure[]{
    \begin{minipage}[t]{0.45\linewidth}
		\centering
		\includegraphics[width=3in,height=3in]{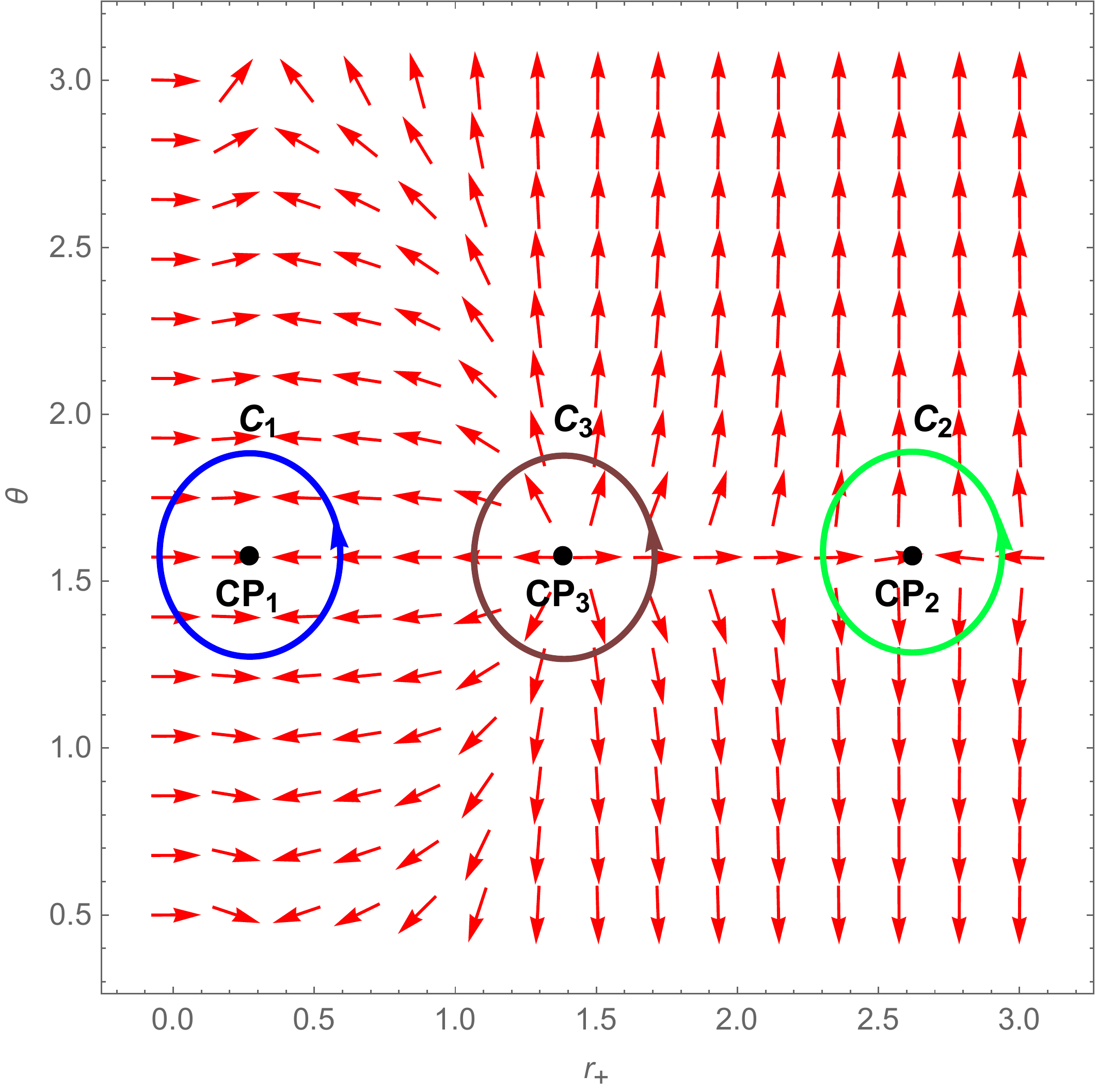}
		\end{minipage}%
            }%
    \subfigure[]{
    \begin{minipage}[t]{0.45\linewidth}
		\centering
		\includegraphics[width=2.6in,height=2in]{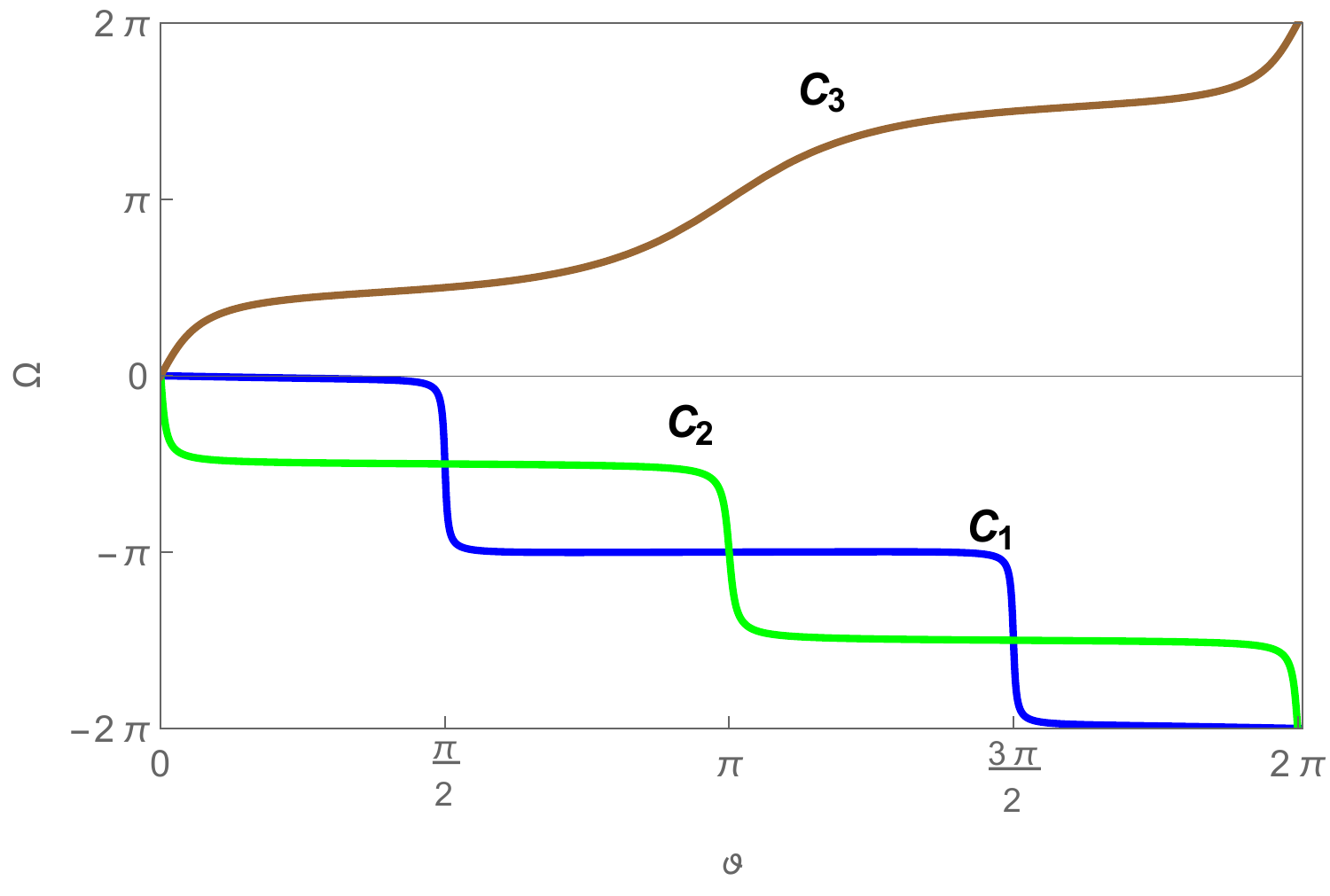}
		\end{minipage}%
            }%
     \centering
     \caption{$d=6$(case 2): (a) The red arrows represent the vector field $n$ for the charged TBHs in the GCE of PMI-massive gravity. The black dots are $(r_+, \theta) =(0.2409, \pi/2 )$, $(r_+, \theta) =(2.6513, \pi/2 )$ and $(r_+, \theta) =(1.3965, \pi/2 )$ which represent the critical points $CP_1$, $CP_2$  and $CP_3$, respectively. (b) The deflection angle $\Omega (\vartheta)$ as a function of $\vartheta$ for contours $C_1$ (blue curve), $C_2$ (green curve) and $C_3$ (brown curve).}
\end{figure}
\subsection{$d= 5$}
According to the analysis in subsection \textbf{A}, when $d\ge5$, the system has two critical points and a reentrant phase transition occurs. This indicates that the system experiences a phase transition characterized by the LBH/SBH/LBH phase transition. let's take $d=5$ as an example. The vector field $n$ in Fig. 2(a) illustrates the existence of two critical points positioned at $(r_+, \theta) =(1.4510, \pi/2)$ and $(r_+, \theta) =(0.89667, \pi/2)$, respectively. The contours $C_1$ and $C_2$ enclose critical points ${CP_1}$ and ${CP_2}$, respectively. $C_3$ is outside the critical points and $C_4$ encloses all two critical points. We can see from Fig. 2(b), the values of the deflection angle $\Omega (2\pi)$ along the contours $C_1$, $C_2$, $C_3$ and $C_4$ are $-2\pi, 2\pi$, 0 and $0$. Those critical points, enclosed by the contours $C_1$ and $C_2$, possess opposite topological charges: $Q_{CP_1}=-1$ and $Q_{CP_2}=1$. Again, $CP_2$ is a novel critical point. The topological charges of the contours $C_3$ and $C_4$ are both zero. Therefore, the total topological charge is $Q=Q_{CP_1}+Q_{CP_2}=0$.
In \cite{ba}, only conventional ${CP_1}$ appears as a critical point in the phase space. Then novel ${CP_2}$ fails to minimize Gibbs free energy. Notably, the absence of any first-order phase transition near ${CP_2}$ supports the findings in \cite{sw1}, which say that novel critical point cannot serve as an indicator of the presence of the first-order phase transition.
\subsection{$d=6$}
\textbf{Case 1:} Similarly, the triple point phenomenon, accompanied by the LBH/IBH/LBH phase transition, becomes apparent in $d\ge 6$. Intriguingly, in such case, three critical points are observed. We simply set $d=6$. Fig. 3(a) illustrates the representation of the vector field $n$ in this context. We construct the contours $C_1$ and $C_2$ to enclose conventional critical points $CP_1$ and $CP_2$, respectively. The contour $C_3$ encloses novel critical point $CP_3$ and contour $C_4$ encloses all three critical points. We can observe that these critical points located at $(r_+, \theta) =(0.7759, \pi/2 )$, $(r_+, \theta) =(2.3588, \pi/2 )$ and $(r_+, \theta) =(1.2157, \pi/2 )$, respectively. Fig. 3(b) represents the behavior of deflection angle for contours $C_1$, $C_2$, $C_3$ and $C_4$. Then, we have two conventional topological charges $Q_{CP_1}=Q_{CP_2}=-1$ and a novel topological charge $Q_{CP_3}=1$. Thus the total topological charge of is $Q=Q_{CP_1}+Q_{CP_2}+Q_{CP_3}=-1$ which is same as the topological charge for the contour $C_4$ as it enclosed all three critical points.
The triple point phenomenon is fascinating, in order for this phenomenon to occur, two of these critical points (conventional critical points ${CP_1}$ and ${CP_2}$) must be physical. However, the third critical point (referred to as the novel critical point $CP_3$) is considered unphysical and incapable of minimizing the Gibbs free energy.

\textbf{Case 2:} For the case 1, there exist two physical critical points. However, by varying at least one parameter, we can obtain three critical points, only one of which is physically meaningful while the other two are non-physical (without minimizing Gibbs free energy). To achieve this, we made adjustment to the nonlinear electromagnetic sector parameter, as indicated in TABLE I. As a result, the system exhibits a vdW type phase transition, which occurs in dimensions $d\ge 6$  as well. Then, we have three critical points located at $(r_+, \theta) =(0.2409, \pi/2 )$, $(r_+, \theta) =(2.6513, \pi/2 )$ and $(r_+, \theta) =(1.3965, \pi/2 )$, respectively. The  normalized vector field $n$ is plotted in Fig. 4(a) in the $r_{+}-\vartheta$ plane. Fig. 4(b) gives $\Omega (\vartheta)=-2\pi, -2\pi$ and $2\pi$ for $\vartheta=2\pi$. Thus, the topological charges associated with the critical points surrounded by the contours $C_1$, $C_2$ and $C_3$ are given by $Q_{CP_1}=-1$, $Q_{CP_2}=-1$, and $Q_{CP_3}=1$ respectively. Consequently, the total topological charge is given by $Q=Q_{CP_1}+Q_{CP_2}+Q_{CP_3}=-1$. Notably, even though $CP_2$ here is a traditional critical point, it doesn't minimize the Gibbs free energy. To resolve this disagreement, let's move on to the next section.
\begin{figure}[htbp]
	\centering
	\subfigure[]{
    \begin{minipage}[t]{0.4\linewidth}
		\centering
		\includegraphics[width=2.6in,height=1.6in]{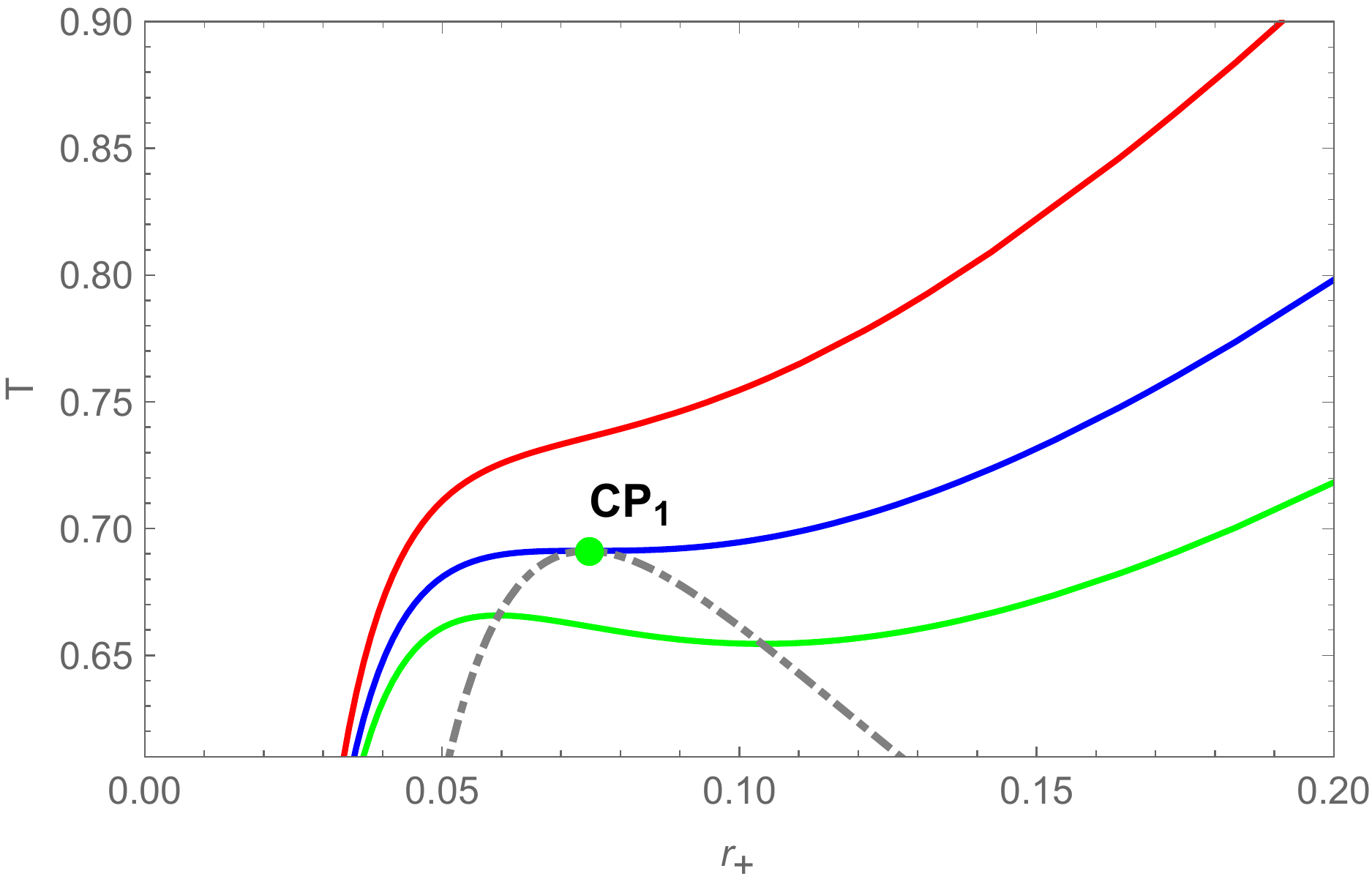}
		\end{minipage}%
            }%
    \subfigure[]{
    \begin{minipage}[t]{0.4\linewidth}
		\centering
		\includegraphics[width=2.6in,height=1.6in]{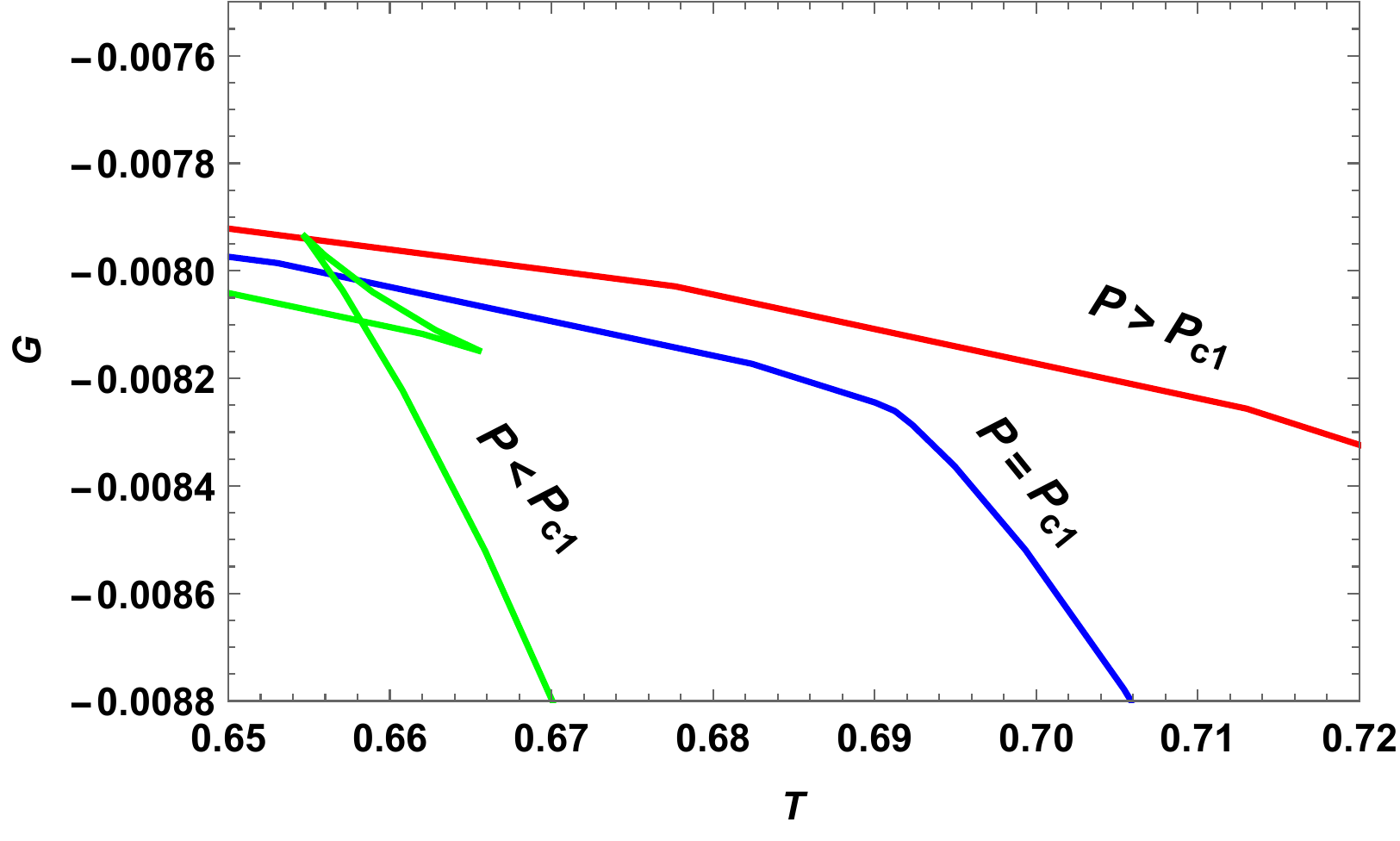}
		\end{minipage}%
            }%
     \centering
     \caption{$d=4$: (a) Isobaric curves (colored solid curves) for the charged TBHs in the GCE of PMI-massive gravity shown in the $T-r_{+}$ plane. The gray dashed curve is for the extremal points of the temperature. The conventional critical point is marked with the green dot. Pressure of the isobars increases from bottom to top. (b) The behavior of Gibbs free energy $G$ near critical point $CP_1$ (with pressure $P_{c1}$).}
\end{figure}
\section{Nature of Critical points}
By utilizing equation of state (9), the colored solid curves and gray dashed curve were employed to depict the behavior of isobaric curves and critical temperature, respectively, as shown in Fig. (5a), Fig. (6a), Fig. (7a) and Fig. (8a), each isobaric curve exhibits distinct extremal points of temperature. Based on our analysis of the topology of BH criticality, we find that the green dots with negative topological charges correspond precisely to the maximum extreme points of temperature. Furthermore, it has been observed that there are unstable regions near these traditional critical points, which can be eliminated through the Maxwell's equal area law. We have to emphasize that the unstable regions here only refer to the intermediate black hole branchs. However, our research also revealed that for the black dot with positive topological charge corresponding to the minimum extreme point of temperature, Maxwell's equal area law is not applicable near such point. These findings have shown that the outcomes we obtained are almost in close agreement with the resolution put forth in \cite{sw1}. This resolution suggests that the existence of a conventional critical point, characterized by a topological charge of -1, can act as an effective indicator for identifying the existence of a first-order phase transition in its surrounding region. However, it is important to note that this observation may not be applicable in all cases. For instance, in the special case of $d= 6$ (referred to as case 2), it has been demonstrated in \cite{ba} that the $P-T$ diagram exclusively displays the presence of the traditional critical point $CP_1$, with no occurrence of a first-order phase transition near the traditional critical point $CP_2$. The investigation and analysis of such phenomena have been discussed in recent research studies \cite{Ye1, NC1}.
\begin{figure}[htbp]
	\centering
	\subfigure[]{
    \begin{minipage}[t]{0.4\linewidth}
		\centering
		\includegraphics[width=2.6in,height=1.6in]{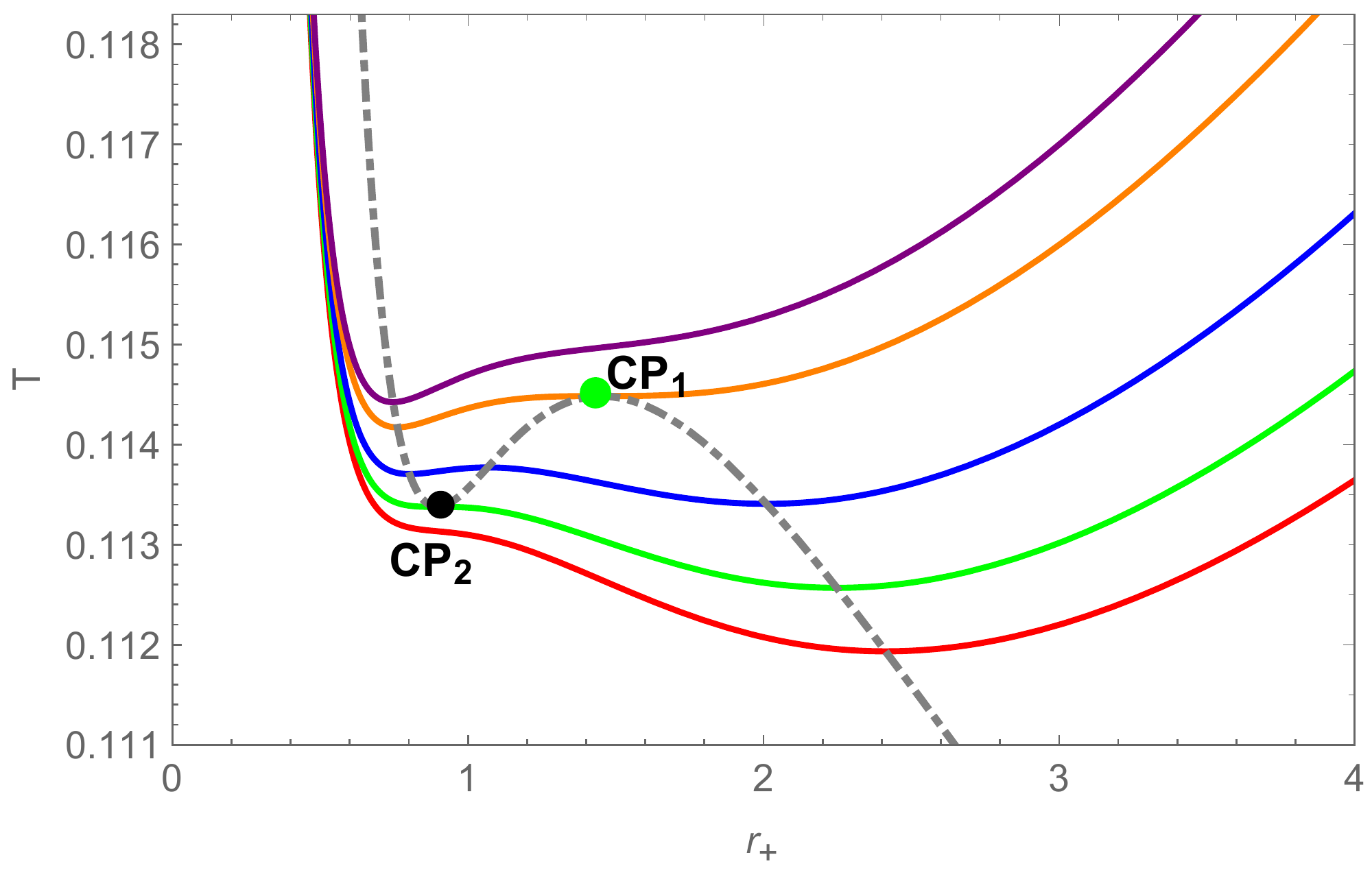}
		\end{minipage}%
            }%
    \subfigure[]{
    \begin{minipage}[t]{0.4\linewidth}
		\centering
		\includegraphics[width=2.6in,height=1.6in]{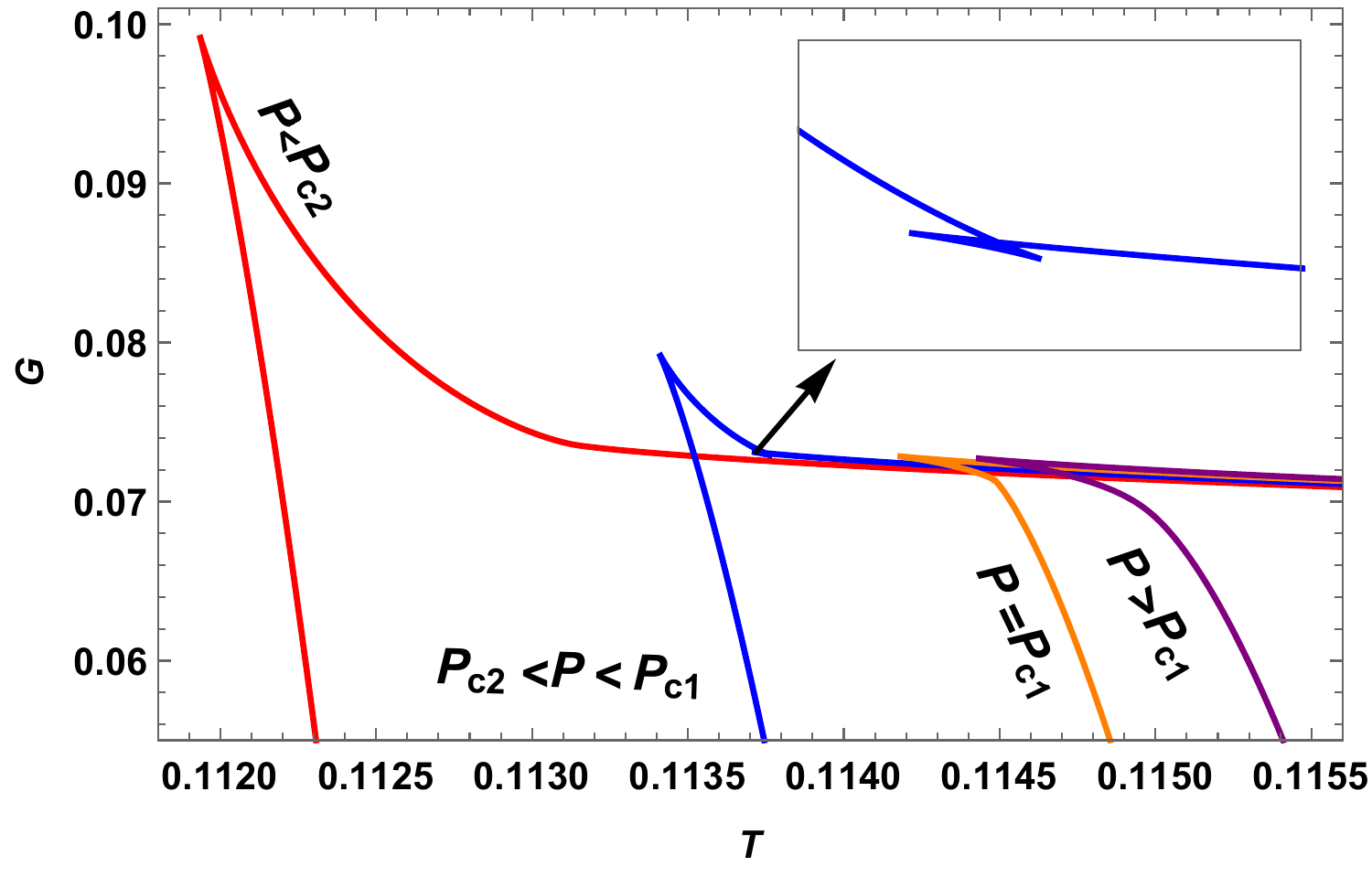}
		\end{minipage}%
            }%
     \centering
     \caption{$d=5$: (a) Isobaric curves (colored solid curves) for the charged TBHs in the GCE of PMI-massive gravity shown in the $T-r_{+}$ plane. The gray dashed curve is for the extremal points of the temperature. The conventional and novel critical points are marked with the green and black dots, respectively. Pressure of the isobars increases from bottom to top. (b) The behavior of Gibbs free energy $G$ near critical points $CP_1$ and $CP_2$ (with pressures $P_{c1}$ and $P_{c2}$, respectively).}
\end{figure}
\begin{figure}[htbp]
	\centering
	\subfigure[]{
    \begin{minipage}[t]{0.4\linewidth}
		\centering
		\includegraphics[width=2.6in,height=1.6in]{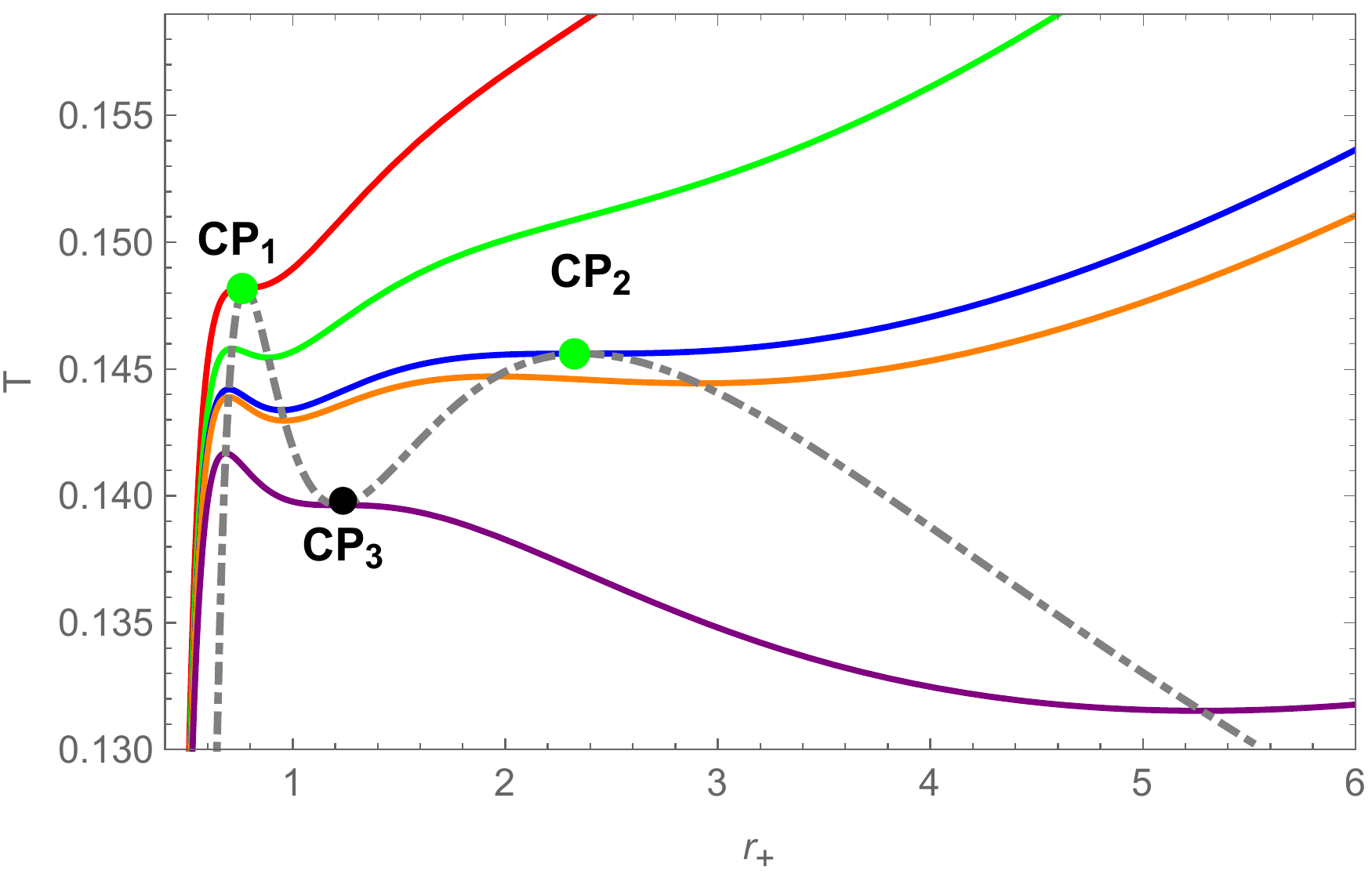}
		\end{minipage}%
            }%
    \subfigure[]{
    \begin{minipage}[t]{0.4\linewidth}
		\centering
		\includegraphics[width=2.6in,height=1.6in]{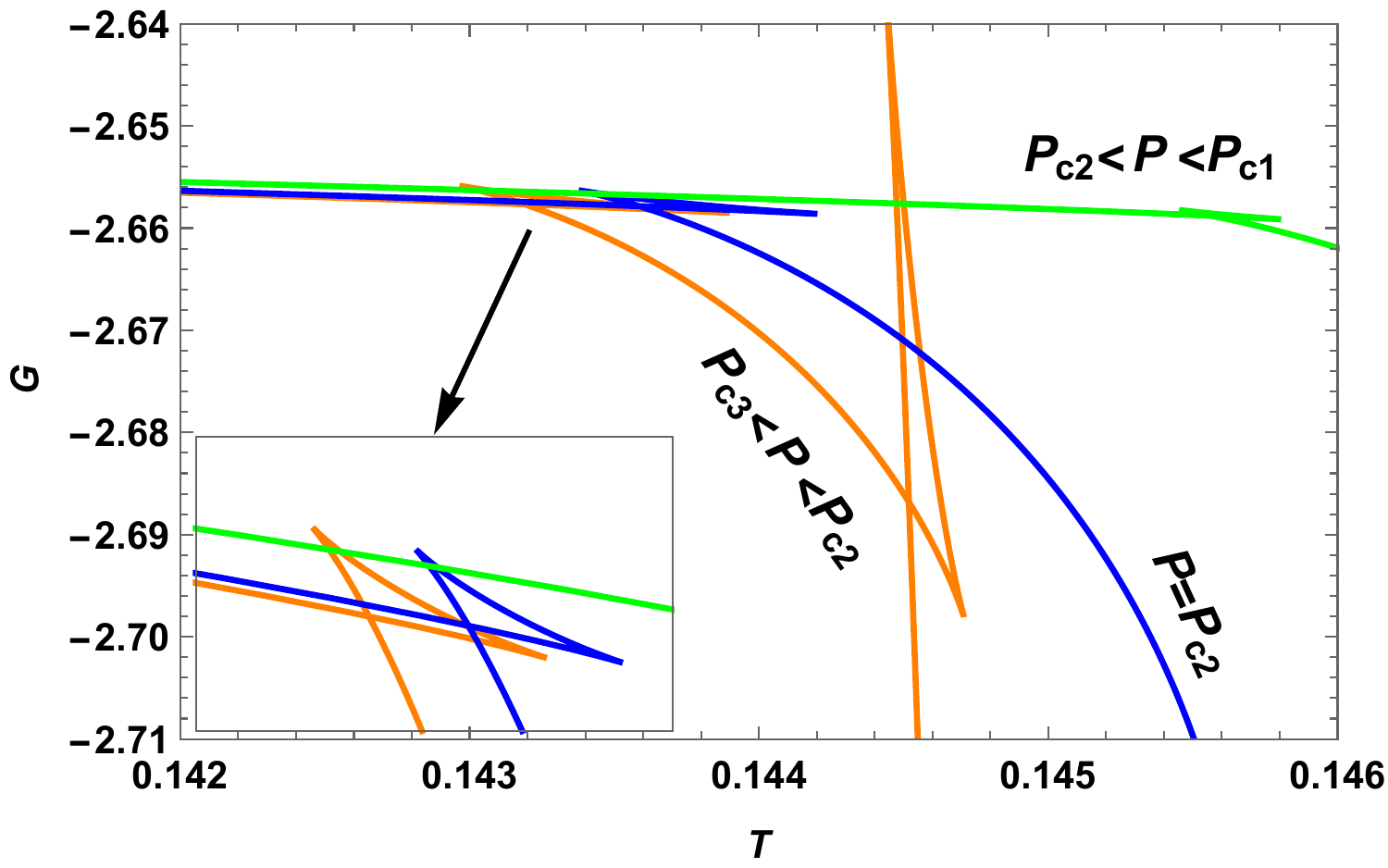}
		\end{minipage}%
            }%
     \centering
     \caption{$d=6$(case 1): (a) Isobaric curves (colored solid curves) for the charged TBHs in the GCE of PMI-massive gravity shown in the $T-r_{+}$ plane. The gray dashed curve is for the extremal points of the temperature. The conventional and novel critical points are marked with the green and black dots, respectively. Pressure of the isobars increases from bottom to top. (b) The behavior of Gibbs free energy $G$ near critical points $CP_1$, $CP_2$ and $CP_3$ (with pressures $P_{c1}$, $P_{c2}$ and $P_{c3}$, respectively).}
\end{figure}
\begin{figure}[htbp]
	\centering
	\subfigure[]{
    \begin{minipage}[t]{0.4\linewidth}
		\centering
		\includegraphics[width=2.6in,height=1.6in]{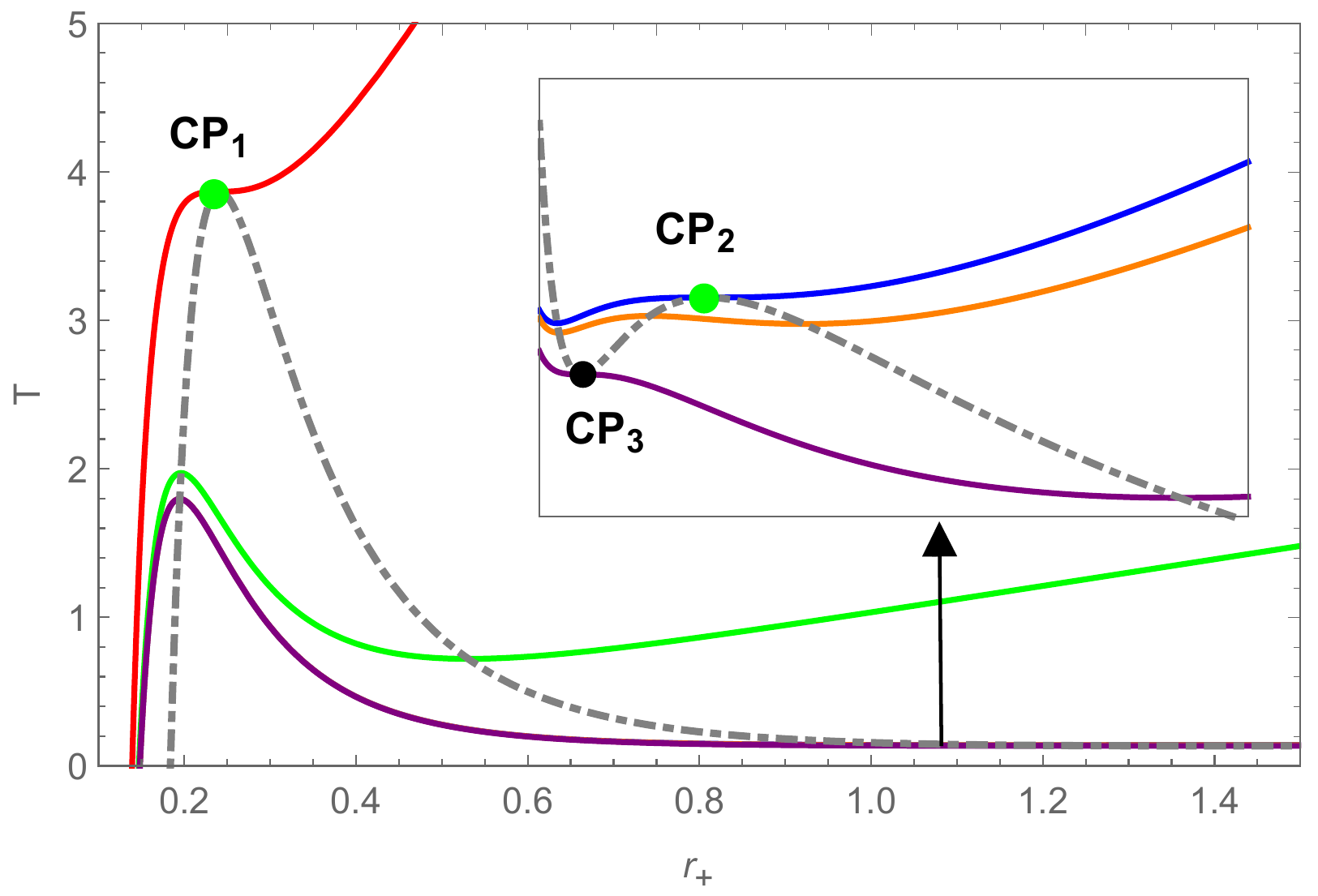}
		\end{minipage}%
            }%
    \subfigure[]{
    \begin{minipage}[t]{0.4\linewidth}
		\centering
		\includegraphics[width=2.6in,height=1.6in]{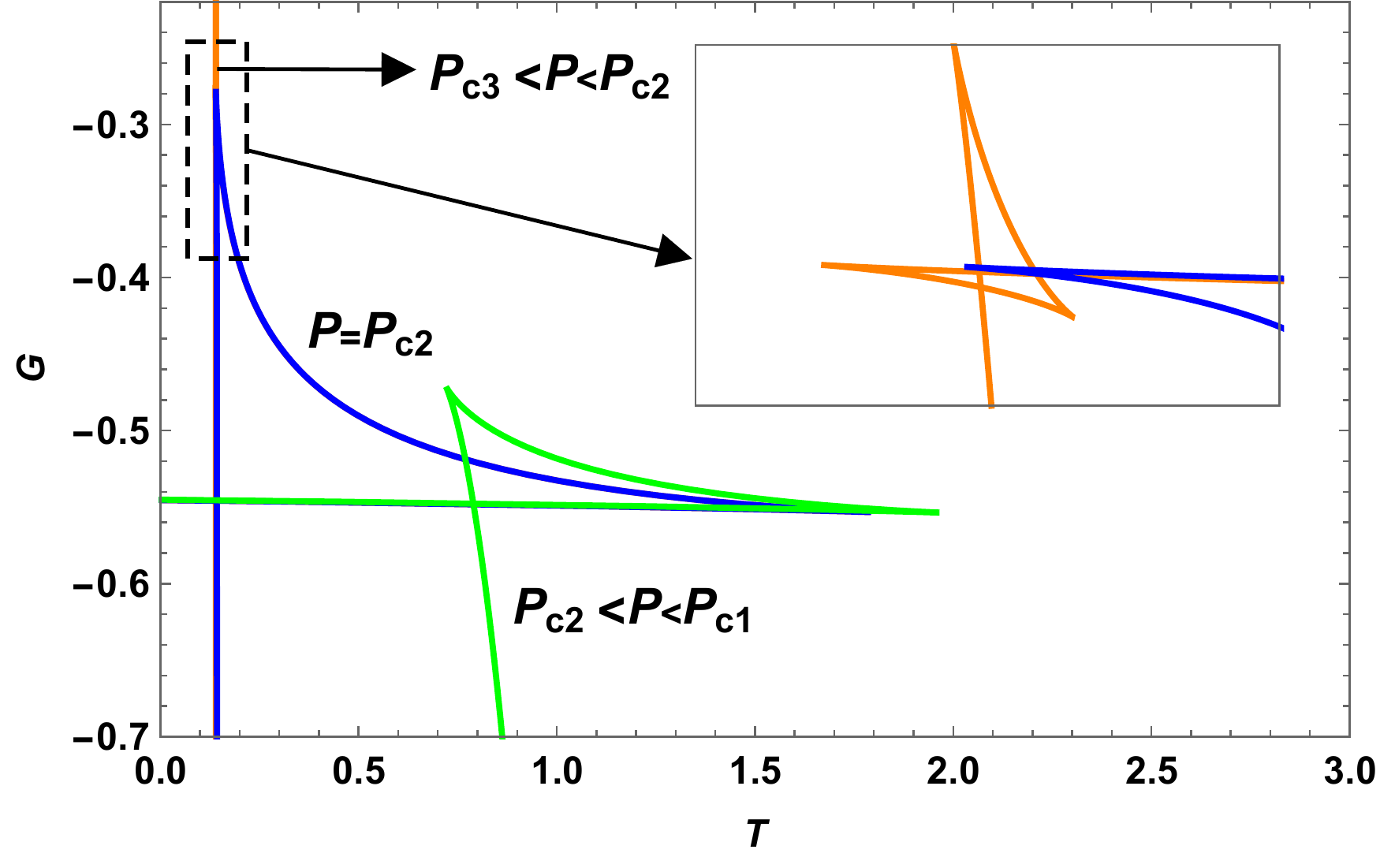}
		\end{minipage}%
            }%
     \centering
     \caption{$d=6$(case 2): (a) Isobaric curves (colored solid curves) for the charged TBHs in the GCE of PMI-massive gravity shown in the $T-r_{+}$ plane. The gray dashed curve is for the extremal points of the temperature. The conventional and novel critical points are marked with the green and black dots, respectively. Pressure of the isobars increases from bottom to top. (b) The behavior of Gibbs free energy $G$ near critical points $CP_1$, $CP_2$ and $CP_3$ (with pressures $P_{c1}$, $P_{c2}$ and $P_{c3}$, respectively).}
\end{figure}
In fact, we find that the critical temperature curve divides each isobaric curve into different number of regions, and we determine the phase structure near the critical point based on this number. For $d= 4$,  there is a critical point ($CP_1$). As the pressure increases, the number of phases present in the isobar diminishes near the conventional critical point, which can be considered as an annihilation point. For $d= 5$, for some black holes' parameters, there are two critical points: the traditional $CP_1$ and novel ($CP_2$). As the pressure increases, Fig. (6a) illustrates that the number of phases (whether stable or unstable) increase at the new critical point and subsequently decrease at the conventional critical point. For $d= 6$(case 1), there are two conventional critical points ($CP_1$, $CP_2$) and a new critical point ($CP_3$). As the pressure increases, it is evident from Fig. 7(a) that the number of phases increase at the novel critical point and then decrease at the conventional critical points. The behaviors of $G$ (the Gibbs free energy) near these critical points are depicted in Fig. (5b), Fig. (6b), and Fig. (7b). For the special case $d= 6$(case 2), we find that above conclusions also held, from Fig.(8b), in the $P_{c3} <P< P_{c2}$ range, there is an anomaly in the standard swallowtail behavior, but no new phase transitions occur. When the pressure increases to $P_{c2}$, the abnormal behavior disappears, so $CP_2$ can be also used as a phase annihilation point.
\begin{figure}[htbp]
	\centering
	\subfigure[]{
    \begin{minipage}[t]{0.45\linewidth}
		\centering
		\includegraphics[width=3in,height=3in]{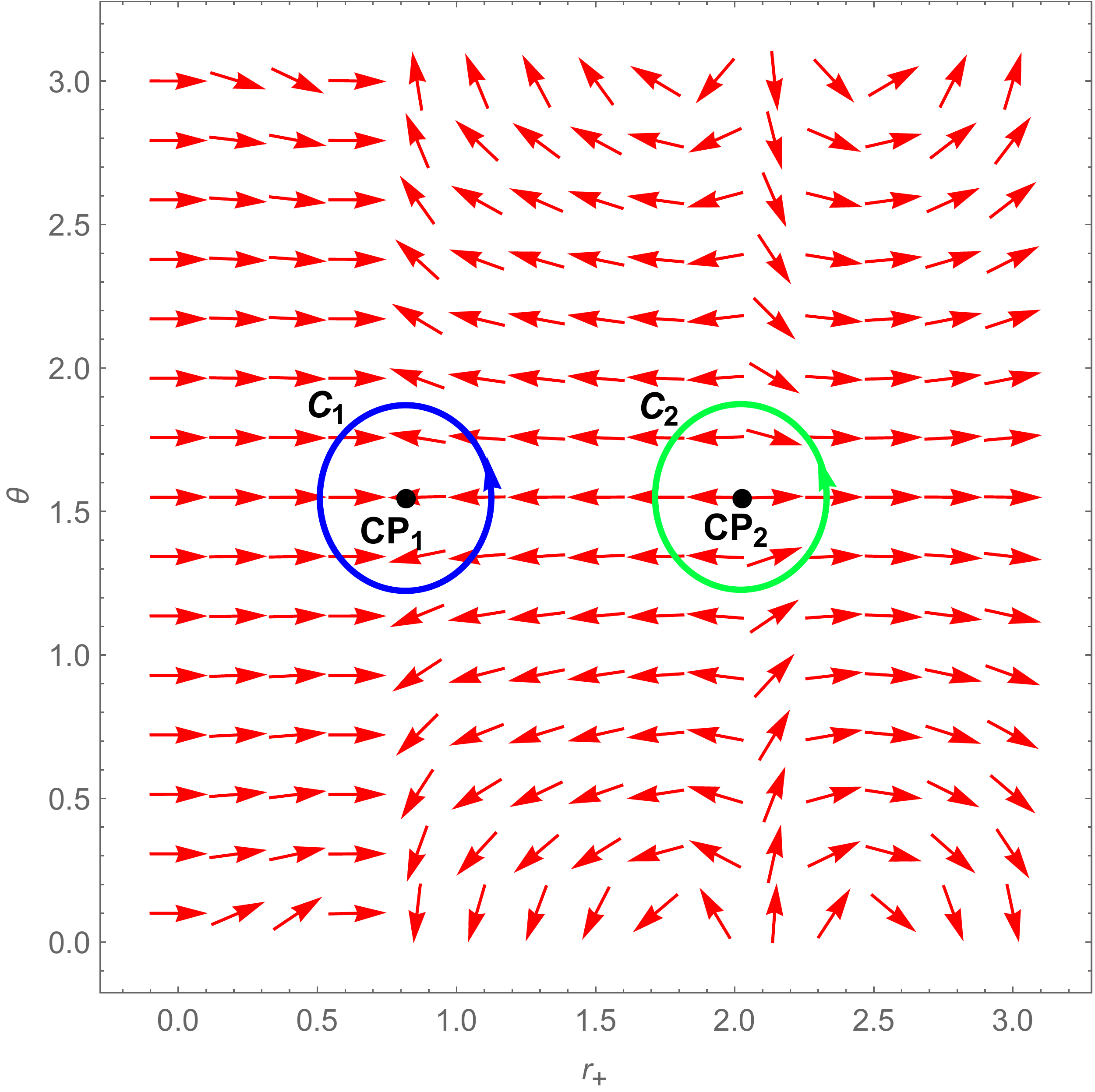}
		\end{minipage}%
            }%
    \subfigure[]{
    \begin{minipage}[t]{0.45\linewidth}
		\centering
		\includegraphics[width=2.6in,height=2in]{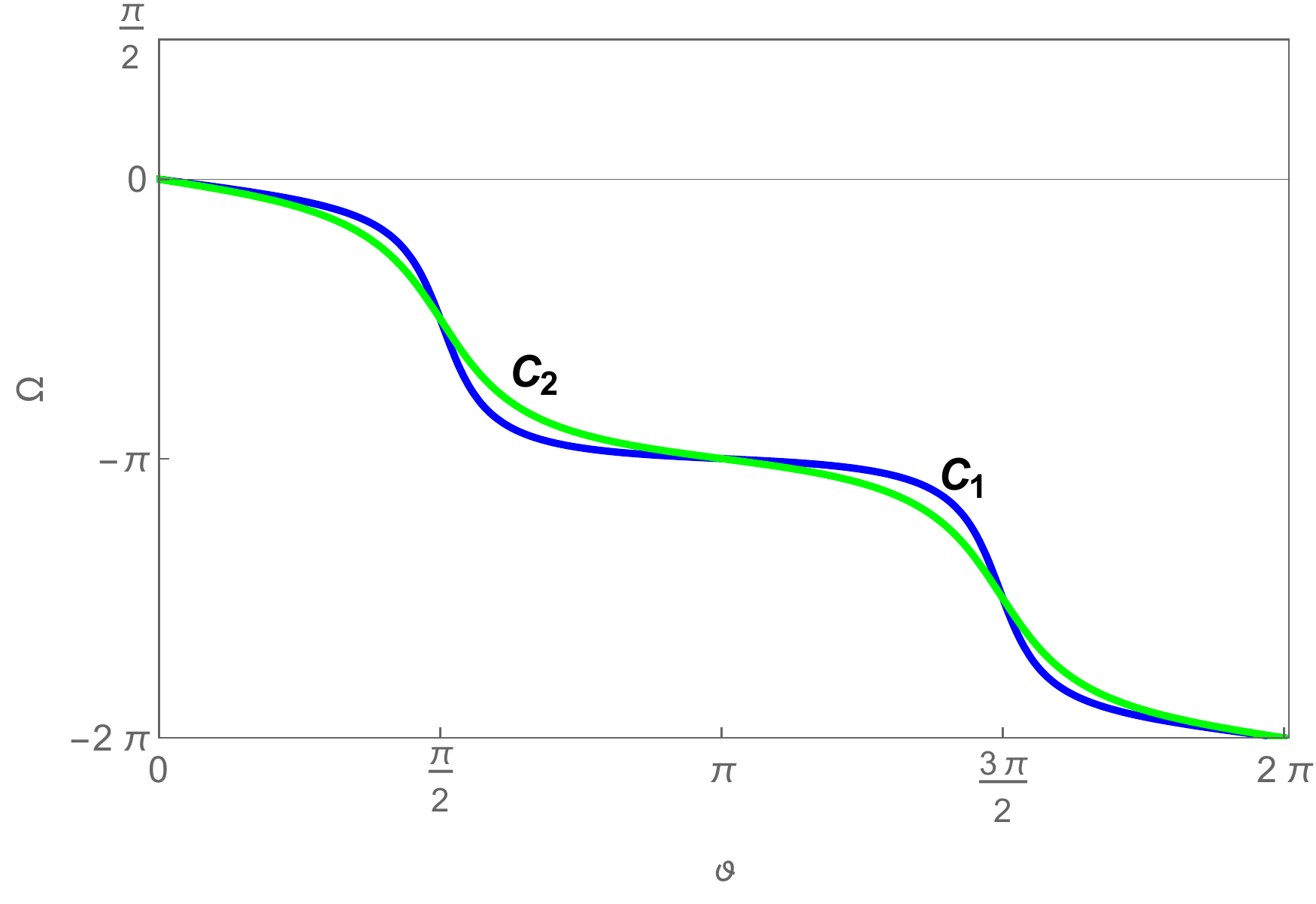}
		\end{minipage}%
            }%
     \centering
     \caption{(a) The red arrows represent the vector field $n$ for the charged TBHs in the canonical ensemble of Maxwell-massive gravity. The black dots from left to right are $(r_+, \theta) =(0.80918, \pi/2 )$ and $(r_+, \theta) =(2.10370, \pi/2 )$ which represent the critical points $CP_1$ and $CP_2$, respectively. (b) The deflection angle $\Omega (\vartheta)$ as a function of $\vartheta$ for contours $C_1$ (blue curve) and $C_2$ (green curve). }
\end{figure}
\section{ Topology under some limiting conditions}
\subsection{ Case-1 $(s=1)$: Topology of Maxwell-massive gravity system}
Next, in the limit $s\to 1$, we would like to investigate the topology of the Maxwell-massive gravity system. The thermodynamic temperature in both canonical ensemble and GCE are given by \cite{SW15}
\begin{equation}
T_{can}=\frac{1}{4 \pi d_2 r_{+}}\left[d_2 d_3 k+\frac{d_1 d_2}{\ell^2} r_{+}^2+m_g^2 \sum_{i=1}^{d_2}\left(c_0^i c_i r_{+}^{2-i} \prod_{j=2}^{i+1} d_j\right)-2q^2r_{+}^{-2d_3} \right],
\end{equation}
and
\begin{equation}
T_{GCE}=\frac{1}{4 \pi d_2 r_{+}}\left[d_2 d_3 k+\frac{d_1 d_2}{\ell^2} r_{+}^2+m_g^2 \sum_{i=1}^{d_2}\left(c_0^i c_i r_{+}^{2-i} \prod_{j=2}^{i+1} d_j\right)-2d_{3}^{2}\Phi ^{2} \right].
\end{equation}
Let's start our discussion with the canonical ensemble. The effective topological factor is $k_{eff}^{(C)}\equiv [k+m_g^2c_0^2c_2]$. Following \cite{SW15}, it is well-established that the standard vdW behavior of a single critical point occurs in $d\ge 4$, thereby yielding a total topological charge of -1. The triple point and anomalous vdW are observed in $d\ge 6$, both of which have three critical points, indicating total topological charge would be also -1. In a 5-dimensional spacetime, two critical radii can be obtained according to the conditions $0<q^2< \frac{3^7}{5\times 2^8}\frac{c_3^4}{c_2^3}$, where $c_2<0$, $c_3>0$. We set $d=5, k=0, m_g=1, c_0=1, c_1=1, c_2=-1.1, c_3=0.8, q=0.4$.  From FIG. 9, we obtain total topological charge is $Q=Q_{CP_1}+Q_{CP_2}=-2$. However, it should be noted that the critical point $CP_2$ yields non-positive definite values for the critical temperature and pressure, making it non-physical. This case provides an example of the parity conjecture of critical points reviewed in \cite{NC1}, which includes negative critical temperature and pressure. A similar situation arises with $d= 6$. This means that the topological class of the TBHs in $d=4,5,6$ remains unchanged. Currently, there is no available data on a reentrant phase transition for $d\ge 7$. Therefore, it cannot be determined whether $d\ge 7$ black holes belong to two different topology classes.

In the case of GCE, the effective topological factor is defined as $k_{eff}^{(GC)}\equiv [k+m_g^2c_0^2c_2-2(d_3/d_2) \Phi^2]$. The standard vdW phase transition (in $d\ge 5$), reentrant phase transition (in $d\ge 6$ ), vdW type behavior (in $d\ge 7$) and triple point phenomena (in $d\ge 7$) are observed \cite{SW15}. These critical phenomena exhibit one, two, and three critical points, respectively. According to the method for calculating the topological charge that described in Sec. III. We find the topological charges of these critical points in one dimension higher have the same values as the GCE of PMI-massive gravity system, indicating that $d=5$ black hole have a total topological charge of -1, whereas the total topological charges of $d\ge 6$ black holes are 0 or -1.
\begin{figure}[htbp]
	\centering
	\subfigure[]{
    \begin{minipage}[t]{0.45\linewidth}
		\centering
		\includegraphics[width=3in,height=3in]{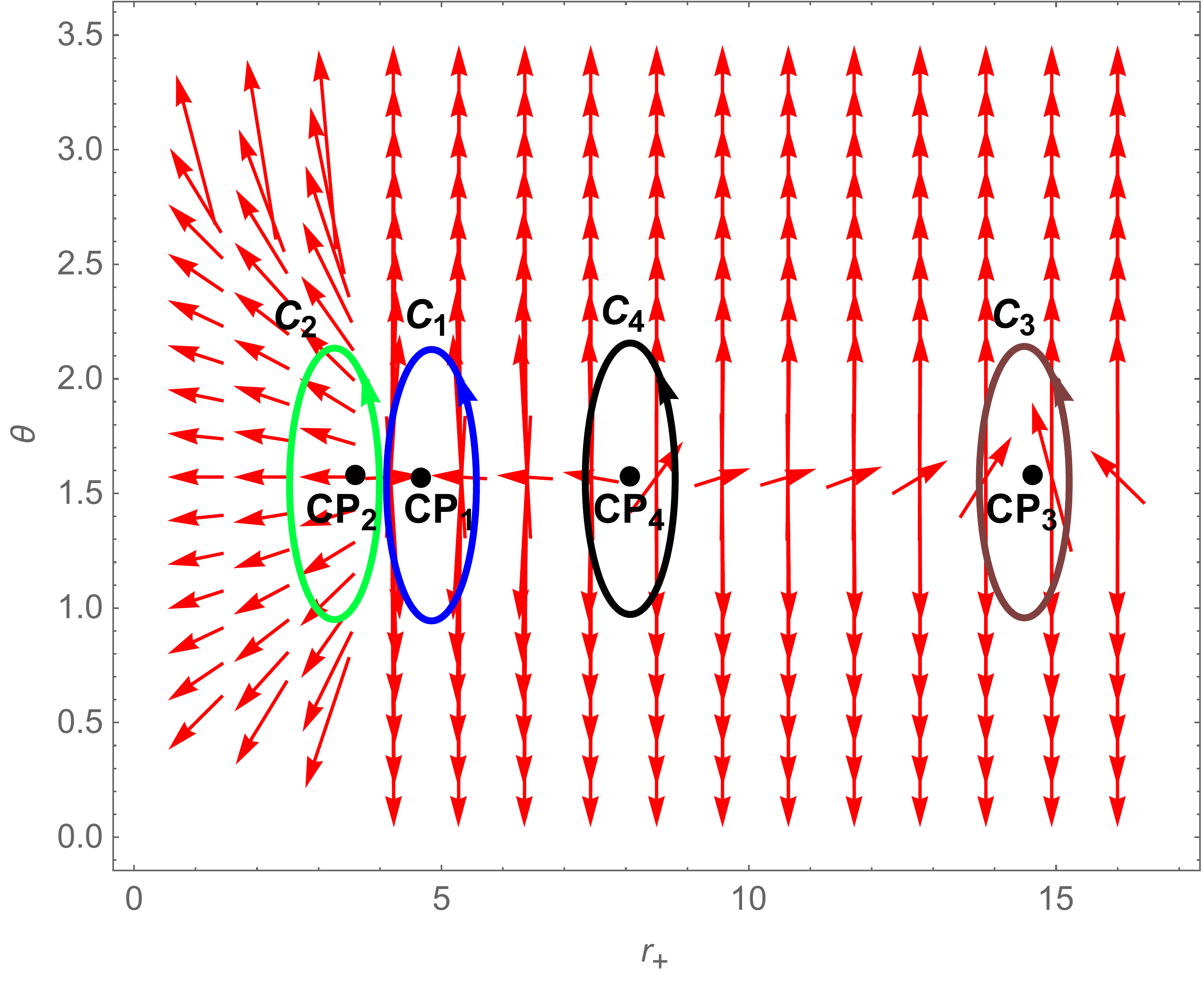}
		\end{minipage}%
            }%
    \subfigure[]{
    \begin{minipage}[t]{0.45\linewidth}
		\centering
		\includegraphics[width=2.6in,height=2in]{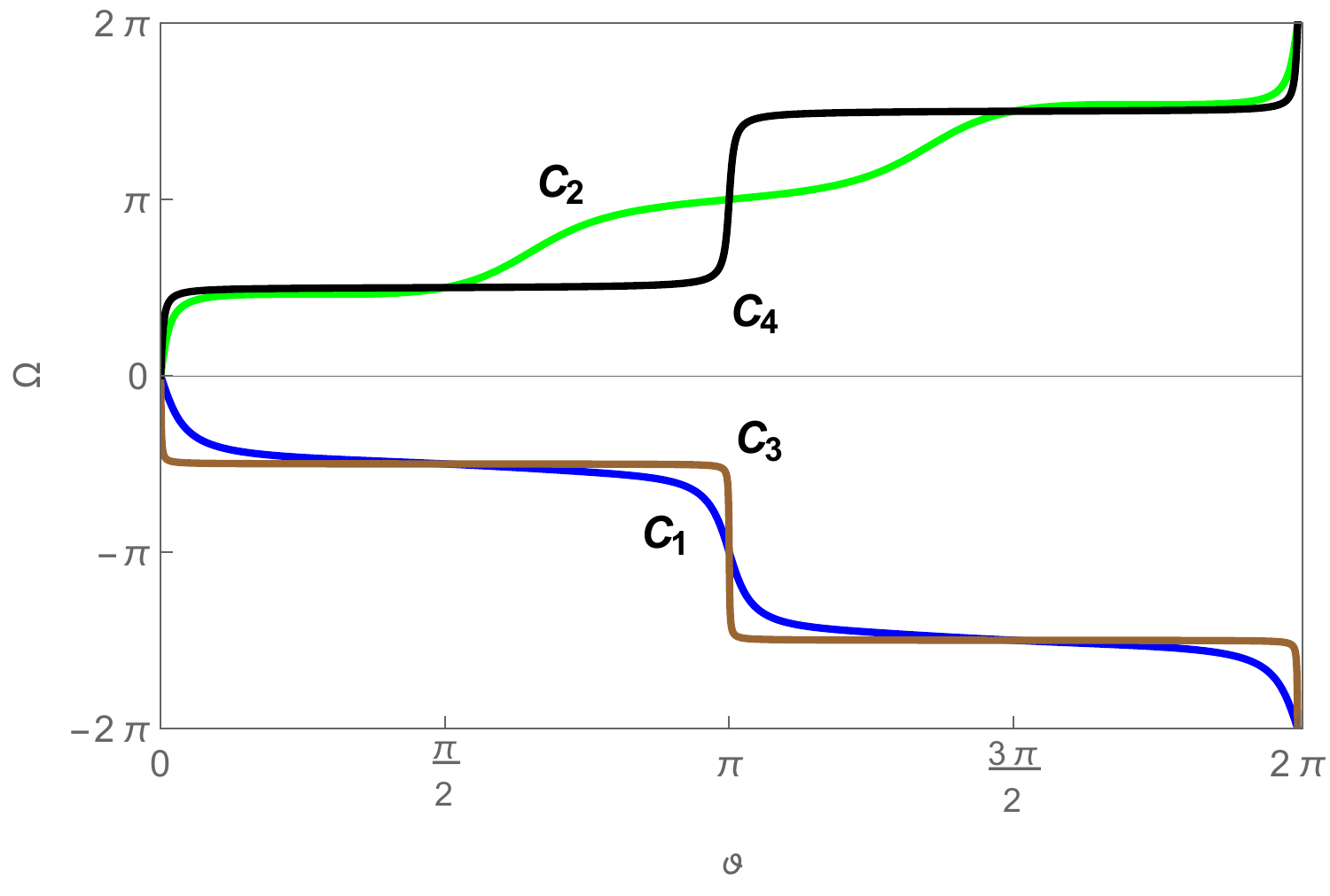}
		\end{minipage}%
            }%
     \centering
     \caption{(a) The red arrows represent the vector field $n$ for the neutral TBHs in pure massive gravity. The critical points $CP_1$, $CP_2$, $CP_3$ and $CP_4$ marked with black dot located at $(r_+, \theta) =(4.70913, \pi/2 )$, $(r_+, \theta) =(3.95971, \pi/2 )$, $(r_+, \theta) =(14.59721, \pi/2 )$ and $(r_+, \theta) =(8.32019, \pi/2 )$. (b) The deflection angle $\Omega (\theta)$ for contours $C_1$ (blue curve), $C_2$ (green curve), $C_3$ (brown curve) and  $C_4$ (black curve).}
\end{figure}
\begin{figure}[htbp]
	\centering
	\subfigure[]{
    \begin{minipage}[t]{0.4\linewidth}
		\centering
		\includegraphics[width=2.6in,height=1.6in]{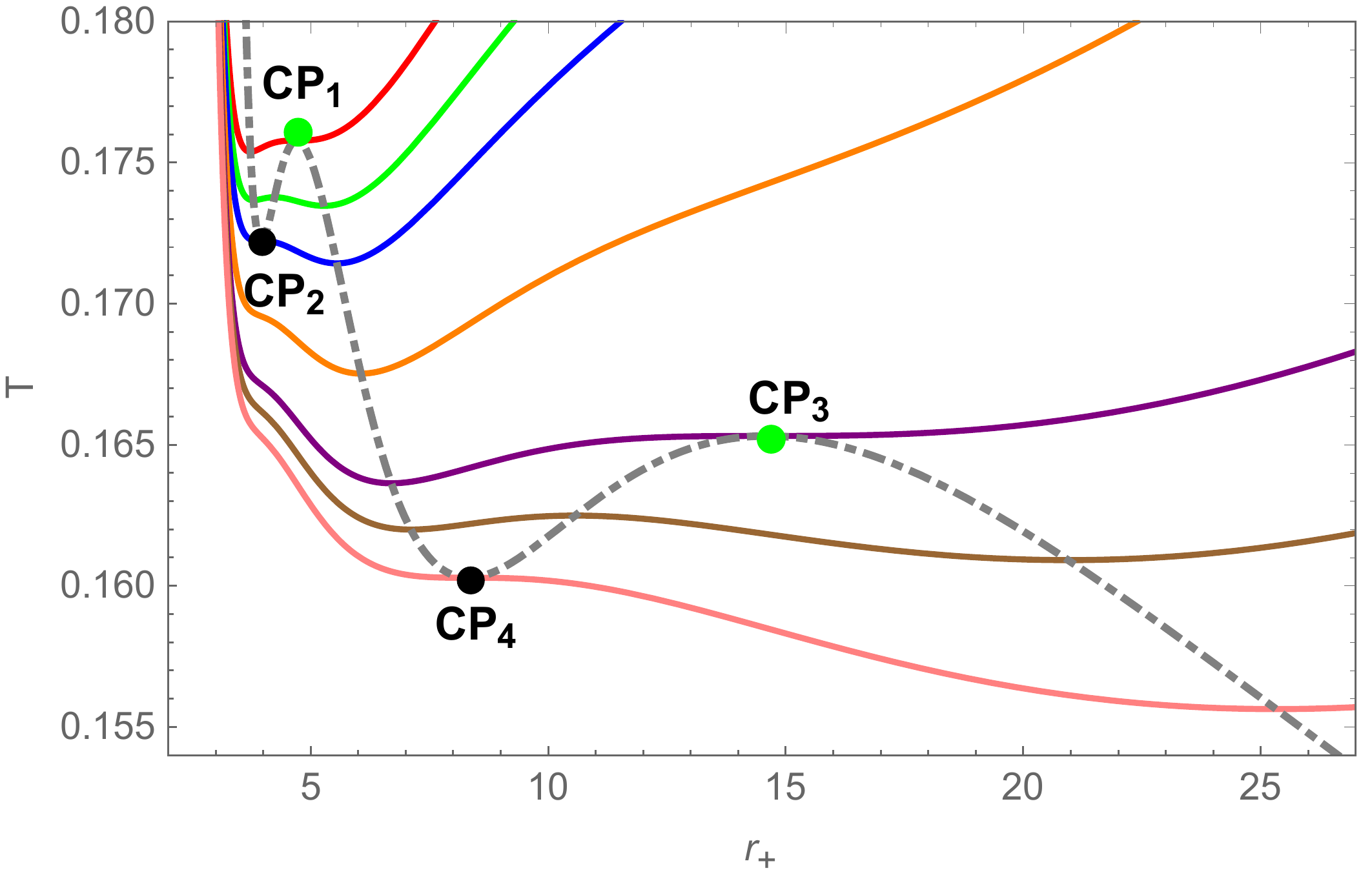}
		\end{minipage}%
            }%
    \subfigure[]{
    \begin{minipage}[t]{0.4\linewidth}
		\centering
		\includegraphics[width=2.6in,height=1.6in]{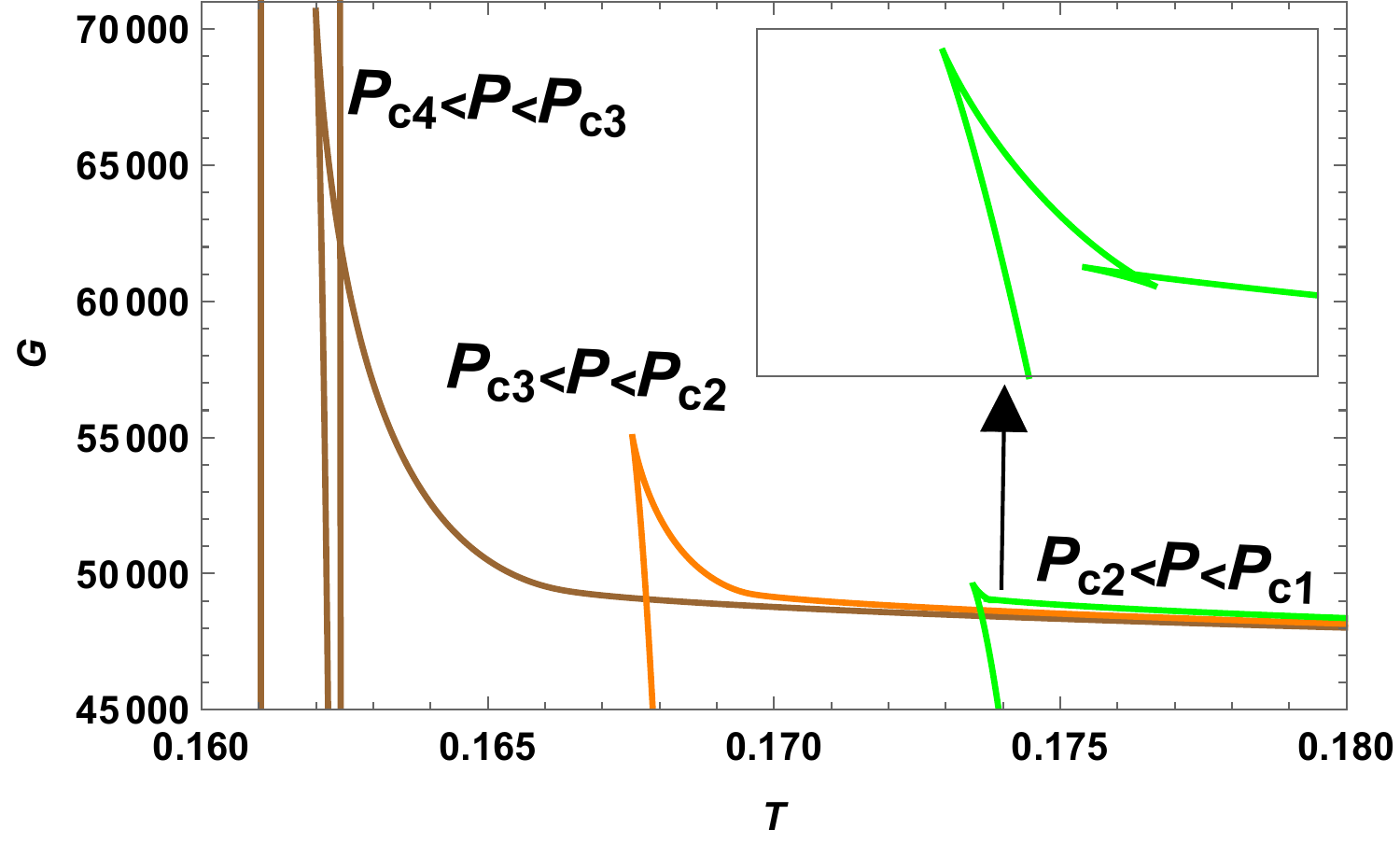}
		\end{minipage}%
            }%
     \centering
     \caption{(a) Isobaric curves (colored solid curves) for the neutral TBHs in pure massive gravity shown in the $T-r_{+}$ plane. The gray dashed curve is for the extremal points of the temperature. The conventional and novel critical points are marked with the green and black dots, respectively. Pressure of the isobars increases from bottom to top. (b) The behavior of Gibbs free energy $G$ near critical points $CP_1$, $CP_2$, $CP_3$ and $CP_4$ (with pressures $P_{c1}$, $P_{c2}$, $P_{c3}$ and $P_{c4}$, respectively).}
\end{figure}
\subsection{ Case-2 $(\Phi=0 ~or ~q=0)$: Topology of pure massive gravity system}
Finally we consider the topology of pure massive gravity system. The thermodynamic temperature of the neutral TBHs is given by \cite{ba1}
\begin{equation}
T_{p}=\frac{1}{4 \pi d_2 r_{+}}\left[d_2 d_3 k+\frac{d_1 d_2}{\ell^2} r_{+}^2+m_g^2 \sum_{i=1}^{d_2}\left(c_0^i c_i r_{+}^{2-i} \prod_{j=2}^{i+1} d_j\right) \right].
\end{equation}
These TBHs can produce up to $n=(d-4)$ critical points for $(d \ne 4)$. The final results of the phase transitions in pure massive gravity system are the same as the Maxwell-massive gravity in fixed potential ensemble with the same dimensions ($d=5,6,7$). For $d\ge 8$, an N-fold reentrant phase transition is observed. We would like to reveal the topology for this new phase structure. We refer to an example with $d=10$ and find four critical points, other parameters were set with $k=0, m_g=1, c_0=1, c_1=1, c_2=3, c_3=-4.3, c_4=2.6, c_5=1.2, c_6=-3.5, c_7=2.6, c_8=0$. Fig. 10(a) shows the behavior of the vector field $n$, which has four critical points located at $(r_c, \theta)$. We construct contours $C_1$, $C_2$, $C_3$ and $C_4$ for $(a, b, r_0)=(0.6, 0.4, 4.70913)$, $(a, b, r_0)=(0.6, 0.4, 3.95971)$, $(a, b, r_0)=(0.6, 0.4, 14.59721)$ and $(a, b, r_0)=(0.6, 0.4, 8.32019)$. The deflection angle $\Omega (\vartheta)$ is calculated from eq. (25). Fig. 10(b) gives $\Omega (2\pi)=-2\pi, 2\pi, -2\pi$ and $2\pi$ for these contours $C_1$, $C_2$, $C_3$ and $C_4$, respectively. Thus the topological charges associated with the four critical points are $Q_{CP_1}=-1$, $Q_{CP_2}=1$, $Q_{CP_3}=-1$ and $Q_{CP_4}=1$. Again, $CP_1$ and $CP_3$ are conventional critical points, while $CP_2$ and $CP_4$ represent novel ones. The total topological charge of pure massive gravity system would be $Q=Q_{CP_1}+Q_{CP_2}+Q_{CP_3}+Q_{CP_4}=0$. As expected, there is no new topology class appears, meaning the topological classification results of the neutral TBHs are the same as the charged TBHs in the GCE of Maxwell-massive gravity. In Fig. 11(a), the critical points ${CP_1}$ and ${CP_3}$ with negative topological charge shown in the green dot correspond to the phase annihilation points; and the novel critical points ${CP_2}$ and ${CP_4}$ by the black dot correspond to the phase creation points. The phase structure near these critical points can also be characterized by the Gibbs free energy $G$, as shown in Fig. 11(b).
\begin{table}[H]
	\centering
	\caption{The total topological charge for different massive gravity theory systems.}
\begin{tabular}{c|l|l|l|l}
 \hline \hline
          Theory                           &Ensemble                  &\makecell[c]{Number of \\ critical Points }           & Dimensions                                         & \makecell[c]{Total topological \\charge }\\
  \hline\multirow{3}{*}{\makecell[c]{PMI-massive~gravity \\$s\ne1$}}&\multirow{3}{*}{GCE}      & 1                           &in $d\ge 4$                                         & -1  \\
                                            &                                   & 2                                          &in $d\ge 5$                                         & 0  \\
                                           &                                    & 3                                          &in $d\ge 6$                                         &-1\\
  \hline&                                   \multirow{3}{*}{Canonical ensemble} & 1                                          &in $d\ge 4$                                         & -1  \\
                                           &                                    & 2(Include negative $T_c$ and $P_c$)        &in $d=5,6$                                          & -2 \\
\multirow{3}{*}{\makecell[c]{Maxwell-massive gravity \\ $s\to 1$  } }&          & 3                                          &in $d\ge 6$                                         & -1  \\\cline{2-5}
                                           & \multirow{3}{*}{GCE}               & 1                                          &in $d\ge 5$                                         & -1  \\
                                           &                                    & 2                                          &in $d\ge 6$                                         & 0 \\
                                           &                                    & 3                                          &in $d\ge 7$                                         & -1 \\
\hline                                     & \multirow{4}{*}{-}                 & 1                                          &in $d\ge 5$                                         & -1  \\
\multirow{3}{*}{\makecell[c]{Pure~massive~gravity \\$\Phi=0 ~or ~q=0 $}}     &  & 2                                          &in $d\ge 6$                                         & 0 \\
                                           &                                    & 3                                          &in $d\ge 7$                                         & -1 \\
                                           &                                    & 4                                          &in $d\ge 8$                                         & 0 \\
\hline \hline
\end{tabular}
\end{table}
\section{CONCLUSIONS} \label{sec4}
By applying Duan's $\phi$-mapping theory, we studied the topology of nonlinearly charged-AdS TBHs thermodynamics. The critical points of PMI-massive gravity system are classified. Under the grand canonical ensemble with fixed potential $\Phi$, we observed a diverse range of phase structures featuring multiple critical points. By considering various dimensions, the phase structure of TBHs can exhibit one, two, or three critical points that correspond to standard vdW behavior in dimensions $d\ge 4$, reentrant phase transitions in dimensions $d\ge 5$, vdW type phase transition in dimensions $d\ge 6$, and even triple point phenomena in dimensions $d\ge 6$. The total topological charges with these critical points are summarized in TABLE III. Our findings indicate that $d=4$ black hole have a total topological charge of -1, which should be classified within the same topology as charged Gauss Bonnet black holes \cite{Ye1}. Whereas $d\ge 5$ black holes possess the total topological charges of 0 or -1, which belong to two different topology classes. These black holes, which in different dimensions, exhibit different phase structures: SBH/LBH, LBH/SBH/LBH, and LBH/IBH/LBH, although the first and third exhibit different phase structures, they may belong to the same topological class within certain parameter spaces (as disscussed in Sec. III). This findings disaccord with the proposal in \cite{NC1} which says that the topological change can be a prognostic indicator of the change in phase structures. In addition, reference \cite{MR1} also draws an inconsistent conclusion that black holes belonging to different topological classes may have the same phase structure.

Based on the topology of BH criticality, we demonstrated that the isobaric curve exhibits two kinds of critical points. The traditional critical point, characterized by negative topological charge, coincides with the maximum extreme point of temperature. In contrast, the novel critical point, featuring positive topological charge, corresponds to the minimum extreme point of temperature. Additionally, the critical temperature curve divides each isobaric curve into different number of phases (stable or unstable). As the pressure increases, the number of phases increases at the new critical point and decreases at the conventional critical point. Our findings provide a support for the updated classification proposed in \cite{Ye1,Ye2}, which distinguishes conventional and novel critical points as phase annihilation and phase creation points, respectively. For $d\ge 6$(case 2), there is an anomaly in the standard swallowtail behavior within the $P_{c3} <P< P_{c2}$ range, but no new phase transitions occurred. As pressure increases to $P_{c2}$, the abnormal behavior disappears. Therefore, conventional $CP_2$ can be used as the critical point where this anomaly disappears.

In addition, we investigated two limiting cases. When $s\to 1$, the Maxwell-massive gravity system being readily recovered. As discussed in Sec. V, we find that black holes in $d=4,5,6$ dimensions belong to the same topological class within the canonical ensemble. In the case of the GCE, the results show that $d=5$ black hole have only one topological class, whereas $d\ge 6$ black holes are in two different topology classes. In summary, for the Maxwell-massive gravity system, different topology classes are only obtained in the GCE and such classes may not exist within the canonical ensemble. However, we must acknowledge that the possibility of different topological class cannot be entirely negated in the canonical ensemble. Interestingly, it has been observed that in the context of the fixed potential ensemble, the topological classes of Maxwell massive gravity at one dimension higher are same as those of PMI-massive gravity. This suggests that the topological classes of the TBHs depend not only on the massive couplings ($c_i$) but also on the nonlinear parameter (s). With the absence of $\Phi$, the neutral TBHs in pure massive gravity have the same topological classes as the charged TBHs in same dimensions in the GCE of Maxwell-massive gravity. The corresponding total topological charges for different massive gravity theory systems are summarized in TABLE III. These results align with the critical point parity conjecture proposed in Ref. \cite{Ye1}, which says that, ``the total topological charge is an odd (even) number for odd (even) number of critical points". In this paper, the critical points include not positive definite critical pressure and temperature.

\section*{Acknowledgments}
We are very grateful to Dr. P. K. Yerra for helpful correspondence. We also acknowledge the anonymous referees for their valuable comments on improving our paper. This work is supported by the National Natural Science Foundation of China (Grant No.12265007) and the Doctoral Foundation of Zunyi Normal University of China (BS [2022] 07, QJJ-[2022]-314).




\end{document}